\PassOptionsToPackage{dvipsnames}{xcolor}
\documentclass[acmsmall,screen,nonacm]{acmart}
\pdfoutput=1

\setcopyright{cc}
\setcctype{by-nc-nd}

\usepackage{booktabs}   %
\usepackage{subcaption} %
\usepackage[dvipsnames]{xcolor}
\usepackage{bcprules}
\usepackage{collcell}
\usepackage{listings}
\usepackage[scaled]{beramono}
\usepackage{lipsum}
\usepackage{mathpartir}

\usepackage{mathtools}
\usepackage{mdframed}
\usepackage{amsmath}
\usepackage{stmaryrd}
\usepackage{varioref}
\usepackage{cleveref}
\usepackage{scalerel}
\usepackage{xspace}
\usepackage{calc}
\usepackage{array,varwidth}
\usepackage{makecell}
\usepackage{diagbox}
\usepackage{float}
\usepackage{xfrac}
\usepackage{enumitem}
\usepackage[skip=2.5pt]{caption}
\usepackage{slashbox}
\usepackage{xifthen}
\usepackage{tikz}
\usepackage{tikz-cd}
\usetikzlibrary{arrows}
\usetikzlibrary{fit}
\usepackage{tikz-layers}
\usepackage{cancel}
\usepackage{bm}
\usepackage[export]{adjustbox}

\usetikzlibrary{arrows, arrows.meta, shapes, positioning}
\tikzset{nodearrow/.style={black, ->, >=myarrow},
myarrow/.tip={Latex[width=1mm, length=1mm]},
tracked/.style={draw=black, fill=blue!10},
locs/.style={draw, circle, tracked, opacity=1},
biglocs/.style={locs, minimum size = 12pt},
shared/.style={fill=yellow!20},
circular locs/.style={locs, fill=purple!10},
shared locs/.style={locs, shared},
lambda/.style={draw, cloud, text centered, cloud puffs=15, aspect=2.5},
untracked lambda/.style={lambda, fill=gray!10, dash pattern=on 5pt off 2pt},
tracked lambda/.style={lambda, tracked},
shared lambda/.style={lambda, shared},
undirected/.style={black, -, >=myarrow},
loop/.style={black, ->, >=myarrow, looseness=8},
arena/.style={draw, rectangle, thick, fill=orange!10, opacity=0.5, inner sep=0.05cm},
bigarena/.style={arena, inner sep=0.2cm},
}

\usetikzlibrary{shadows}
\usetikzlibrary{shadows.blur}
\usetikzlibrary{decorations.pathreplacing} %
\tikzcdset{arrow style=tikz,
           diagrams={>=stealth}
         }    %
\tikzcdset{crossing over clearance=.75ex}
\tikzset{hardd/.style={teal,thick}} %
\tikzset{softd/.style={teal, dashed,thick}} %
\tikzset{datad/.style={black,densely dotted}} %
\tikzset{alld/.style={black,thick}} %
\tikzset{expnode/.style={
    asymmetrical rectangle,rounded corners,draw,fill=white,inner sep=2.5pt,
    font=\footnotesize, %
  }} %
\tikzset{blknode/.style={
    asymmetrical rectangle,sharp corners,draw,violet,fill=violet!10,
    font=\footnotesize %
  }} %
\tikzset{dummynode/.style={fill=none,draw=none,asymmetrical rectangle} %
} %
\tikzset{nmute/.style={opacity=.4,blur shadow={shadow opacity=.4}}} %
\tikzset{expn/.style={
    circle,fill=white,draw,minimum size=5pt,inner sep=0pt,outer sep=0pt,
    font=\footnotesize, %
    label={[font=\footnotesize]left:#1}
  }} %
\tikzset{expn/.default={}}
\tikzset{mexpn/.style={opacity=.4,
    circle,fill=white,draw,minimum size=5pt,inner sep=0pt,outer sep=0pt,
    font=\footnotesize, %
    label={[font=\footnotesize,opacity=.4]left:#1}
  }} %
\tikzset{mexpn/.default={}}
\tikzset{rmexpn/.style={opacity=.4,
    circle,fill=white,draw,minimum size=5pt,inner sep=0pt,outer sep=0pt,
    font=\footnotesize, %
    label={[font=\footnotesize,opacity=.4]right:#1}
  }} %
\tikzset{rmexpn/.default={}}
\tikzset{mexpndummy/.style={opacity=.4,
    circle,fill=white,minimum size=5pt,inner sep=0pt,outer sep=0pt,
    label={[font=\footnotesize,opacity=.4]left:#1}
  }} %
\tikzset{mexpndummy/.default={}}

\usepackage{listings,array,varwidth}
\usepackage{mathpartir}
\usepackage{mathtools}
\usepackage{makecell}

\usepackage{diagbox}
\usepackage{float}
\usepackage{varioref}
\usepackage{xfrac}
\usepackage{enumitem}
\usepackage[skip=2.5pt]{caption}
\usepackage{slashbox}

\usepackage{xifthen}

\usepackage{tikz}
\usepackage{xsavebox}
\usetikzlibrary{arrows}

\makeatletter %
\def\arcr{\@arraycr}
\makeatother

\newcommand{\showDOI}[1]{\unskip}

\newtheorem{innercustomgeneric}{\customgenericname}
\providecommand{\customgenericname}{}
\newcommand{\newcustomtheorem}[2]{%
  \newenvironment{#1}[1]
  {%
   \renewcommand\customgenericname{#2}%
   \renewcommand\theinnercustomgeneric{##1}%
   \innercustomgeneric
  }
  {\endinnercustomgeneric}
}
\newcustomtheorem{re-definition}{Definition}
\newcustomtheorem{re-lemma}{Lemma}

\newcommand{\Specsharp}{%
	{\settoheight{\dimen0}{C}Spec\kern-.05em \resizebox{!}{\dimen0}{\raisebox{\depth}{\#}}}}
\newcommand{\Csharp}{%
	{\settoheight{\dimen0}{C}C\kern-.05em \resizebox{!}{\dimen0}{\raisebox{\depth}{\#}}}}

\newcommand{\fun}[1]{\operatorname{#1}}
\newcommand{\DOM}{\fun{dom}}
\newcommand{\DOML}{\fun{dom\ell}}

\definecolor{blue-violet}{rgb}{0.54, 0.17, 0.89}
\definecolor{dark-cyan}{HTML}{135579}
\definecolor{magenta}{HTML}{a8264f}

\lstdefinelanguage{Neutral}%
{morekeywords={abstract,%
  case,catch,char,class,%
  def,else,extends,final,finally,for,%
  if,import,implicit,%
  match,module,%
  new,null,%
  object,override,%
  package,private,protected,public,%
  for,public,return,super,%
  this,trait,try,type,%
  val,var,%
  while,%
  yield,%
  let,end,%
	in,fun,alloc,inc,
  at,withRef,%
  },%
  mathescape=true,%
  sensitive,%
  keywordstyle={\color{black}\bf\ttfamily},%
  commentstyle=\color{OliveGreen},%
  escapebegin=\color{OliveGreen},
  morecomment=[l]//,%
  morecomment=[s]{/*}{*/},%
  morecomment=[s][\color{darkgray}]{@}{\ },%
  morestring=[b]",%
  morestring=[b]',%
  showstringspaces=false%
}[keywords,comments,strings]%

\lstdefinelanguage{OOPSLA21}%
{morekeywords={abstract,%
  case,catch,char,class,%
  def,else,extends,final,finally,for,%
  if,import,implicit,%
  match,module,%
  new,null,%
  object,override,%
  package,private,protected,public,%
  for,public,return,super,%
  this,throw,trait,try,type,%
  val,var,%
  with,while,%
  yield,%
  let,end,%
	in,fun,alloc,inc,
  at,withRef,%
  },%
  mathescape=true,%
  sensitive,%
  keywordstyle={\color{magenta}\bf\ttfamily},%
  commentstyle=\color{magenta},%
  escapebegin=\color{magenta},
  morecomment=[l]//,%
  morecomment=[s]{/*}{*/},%
  morecomment=[s][\color{magenta}]{@}{\ },%
  morestring=[b]",%
  morestring=[b]',%
  showstringspaces=false%
}[keywords,comments,strings]%

\lstdefinelanguage{PolyRT}%
{morekeywords={abstract,%
  case,catch,char,class,%
  def,else,extends,final,finally,for,%
  if,import,implicit,%
  match,module,%
  new,null,%
  object,override,%
  package,private,protected,public,%
  for,public,return,super,%
  this,throw,trait,try,type,%
  val,var,%
  while,%
  yield,%
  let,end,%
	in,fun,alloc,%
  at,withRef,scoped%
  },%
  mathescape=true,%
  sensitive,%
  keywordstyle={\color{dark-cyan}\bf\ttfamily},%
  commentstyle=\color{dark-cyan},%
  escapebegin=\color{dark-cyan},%
  morecomment=[l]//,%
  morecomment=[s]{/*}{*/},%
  morecomment=[s][\color{dark-cyan}]{@}{\ },%
  morestring=[b]",%
  morestring=[b]',%
  showstringspaces=false%
}[keywords,comments,strings]%

\newcommand{\trackvar}[1]{^{\texttt{#1}}}
\newcommand{\trackvarm}[1]{^{\texttt{\ensuremath{#1}}}}
\newcommand{\trackfresh}{^{\texttt{\ensuremath\vardiamondsuit}}}

\newcommand{\weilang}{\ensuremath{\lambda^{\vardiamondsuit}}\xspace}
\newcommand{\cycliclang}{\ensuremath{\lambda_{\circ}}\xspace}
\newcommand{\weipolylang}{\ensuremath{\mathsf{F}_{<:}^{\vardiamondsuit}}\xspace}
\newcommand{\arenalang}{\ensuremath{\mathsf{A}_{<:}^{\vardiamondsuit}}\xspace}

\newcommand{\withlang}{\ensuremath{{\left\{\mathsf{A}\right\}}_{<:}^{\vardiamondsuit}}\xspace}

\makeatletter
\def\ifenv#1{
   \def\@tempa{#1}%
   \ifx\@tempa\@currenvir
      \expandafter\@firstoftwo
    \else
      \expandafter\@secondoftwo
   \fi
}
\makeatother

\newcommand{\Type}[1]{\ensuremath{\ifenv{lstlisting}{\texttt{#1}}{\mathsf{#1}}}}

\newcommand{\ty}[2][]{\ensuremath{\ifthenelse{\isempty{#1}}{#2}{#2^{\,#1}}}}

\newcommand{\Var}{\Type{Var}}

\newcommand{\TRef}{\Type{Ref}}
\newcommand{\TTop}{\Type{Top}}
\newcommand{\TUnit}{\Type{Unit}}
\newcommand{\Loc}{\Type{Loc}}
\newcommand{\Offset}{\Type{Offset}}

\newcommand{\tunit}{\Type{unit}}

\newcommand{\tref}{\Type{ref}}
\newcommand{\trefat}[2]{\mathsf{ref}~#1~\mathsf{at}~#2}
\newcommand{\twithr}[3][]{\ifthenelse{\isempty{#1}}{\mathsf{ref}~#2~\mathsf{as}~x~\mathsf{in}~#3}{\mathsf{ref}~#2~\mathsf{as}~#1~\mathsf{in}~#3}}
\newcommand{\twithc}[2]{\mathsf{local}~#1~\mathsf{in}~#2}

\newcommand{\LC}[1]{\mathsf{LC}(#1)}
\newcommand{\None}{\mathsf{None}}
\newcommand{\WT}[1]{\ensuremath{#1\ \mathsf{wf}}}

\newcommand{\killsep}[2]{\ensuremath{#1 \cap #2 = \varnothing}}
\newcommand{\envto}[1]{\leadsto_{#1}}

\newcommand{\transform}[1]{\llbracket #1 \rrbracket}
\newcommand{\sourcecode}[1]{{\color{dark-cyan}#1}}
\newcommand{\targetcode}[1]{\ensuremath{{\color{magenta} #1 } }}
\newcommand{\transformsource}[1]{\transform{$ \sourcecode{\lstinline|#1|} $ }}
\newcommand{\continuation}{\mathscr{k}}
\newcommand{\continuationof}[2][]{\ifthenelse{\isempty{#1}}{\color{black}{\continuation[}#2\color{black}{]}}{\color{black}{\continuation_{#1}[}#2\color{black}{]}}}

\newcommand{\cpsfresh}[1]{\ensuremath{\underline{#1}}}

\newcommand{\lo}[2]{\ensuremath{#1,#2}}
\newcommand{\lot}[2]{\ensuremath{(\lo{#1}{#2})}}

\newcommand{\TAll}[6]{\ensuremath{\forall f(\ty[#2]{#1} <: \ty[#4]{#3}) . \ty[#6]{#5}}}
\newcommand{\TLam}[3]{\ensuremath{\Lambda f(\ty[#2]{#1}) . {#3}}}
\newcommand{\TApp}[3]{\ensuremath{{#1}\ [ \ty[#3]{#2} ] } }

\newcommand{\ts}[1][]{\ensuremath{\ifthenelse{\isempty{#1}}{\,\vdash\,}{\,\vdash^{\,#1}\,}}}

\newcommand{\flt}{\ensuremath{\varphi}}

\newcommand{\cx}[2][]{\ensuremath{\ifthenelse{\isempty{#1}}{#2}{#2^{\,#1}}}}

\providecommand{\G}{G} %
\renewcommand{\G}[1][]{\cx[#1]{\Gamma}}

\newcommand{\HLBox}[2][teal!12]{\ensuremath{\mathchoice%
  {\setlength{\fboxsep}{.5ex}\colorbox{#1}{$\displaystyle#2$}}%
  {\setlength{\fboxsep}{.5ex}\colorbox{#1}{$\textstyle#2$}}%
  {\setlength{\fboxsep}{.5ex}\colorbox{#1}{$\scriptstyle#2$}}%
  {\setlength{\fboxsep}{.5ex}\colorbox{#1}{$\scriptscriptstyle#2$}}}}%

\newcommand{\QFresh}{\ensuremath{\vardiamondsuit}}
\newcommand{\qbot}{\ensuremath{\varnothing}}
\newcommand{\qfresh}{\ensuremath{\vardiamondsuit}}
\newcommand{\subq}{\ensuremath{\subseteq}}
\newcommand{\qlub}{\ensuremath{\cup}}
\newcommand{\qglb}{\ensuremath{\cap}}
\xsavebox*{SMALLSTAR}{\(\raisebox{.25ex}{\(\qfresh\)}\)}
\newcommand{\starred}[1]{\ensuremath{\mathord{\scalerel*{\xusebox{SMALLSTAR}}{*}}#1}}
\xsavebox*{OVRLP}{$\raisebox{.37ex}[0pt][0pt]{$\mathrlap{\hspace{.415ex}\scaleobj{.5}{\vardiamondsuit}}$}\cap$}
\def\overlap{\ensuremath{\mathbin{\scalerel*{\xusebox{OVRLP}}{\sqcap}}}}

\newcommand{\WF}[1]{\ensuremath{#1\ \mathsf{ok}}}

\newcommand{\norm}[1]{\lvert #1 \rvert}

\newcommand{\qtrans}[2][]{\ensuremath{\ifthenelse{\isempty{#1}}{#2\mathord{*}}{{#2}^{#1}}}}

\newcommand{\BOX}[1]{\fbox{$\strut #1$}}

\newcommand{\FV}{\ensuremath{\operatorname{fv}}}
\newcommand{\FTV}{\ensuremath{\operatorname{ftv}}}

\newcounter{typerule}
\crefformat{typerule}{#2\textit{#1}#3}
\Crefformat{typerule}{#2\textit{#1}#3}

\newcommand{\hole}[1]{\ensuremath{[\,#1\,]}}
\newcommand{\CX}[3][black]{\ensuremath{{\color{#1}#2\ifthenelse{\isempty{#3}}{}{\hole{{\color{black}#3}}}}}}

\newcommand{\ie}{{\em i.e.}\xspace}

\newcommand{\ruledef}[1]{\hypertarget{rule:#1}{\expandafter\csname NoLink#1\endcsname}}
\newcommand{\newrulename}[2]{\expandafter\newcommand\csname NoLink#1\endcsname{#2}}
\newcommand{\rulelab}[1]{\hyperlink{rule:#1}{\expandafter\csname NoLink#1\endcsname}}
\newcommand{\rulefmt}[1]{(\textsc{#1})} %
\newcommand{\rulename}[1]{\rulefmt{\rulelab{#1}}}

\newcommand{\Fsub}{\ensuremath{\mathsf{F}_{<:}}\xspace}

\newcommand{\arenacontent}[1]{\HLBox[pink!60]{\ensuremath{#1}}}
\newcommand{\scopedynamics}[1]{\HLBox[violet!30]{\ensuremath{#1}}}
\newcommand{\scopedynamicstext}[1]{\colorbox{violet!30}{#1}}

\lstdefinelanguage{DOT}%
{morekeywords={val,new},%
  sensitive,%
  morecomment=[l]//,%
  morecomment=[s]{/*}{*/},%
  morestring=[b]",%
  morestring=[b]',%
  showstringspaces=false%
}[keywords,comments,strings]%

\newlength{\trulemargin}
\newlength{\trulewidth}
\newlength{\srulewidth}
\newenvironment{trules}{$\vspace{0.5em}\ba{p{\trulemargin}@{~}p{\trulewidth}@{~}p{\trulemargin}}}{\ea$}
\newenvironment{srules}{$\vspace{0.5em}\ba{p{\trulemargin}@{~}p{\srulewidth}}}{\ea$}

\newcommand{\ba}{\begin{array}}
\newcommand{\ea}{\end{array}}

\newcommand{\ei}{\end{array}}
\newcommand{\bcases}{\left\{\begin{array}{ll}}
\newcommand{\ecases}{\end{array}\right.}

\newcommand{\eg}{{\em e.g.}\xspace}

\newcommand{\dom}{\mbox{\sl dom}}

\newcommand{\judgement}[2]{{\textsf{\textbf{#1}}} \hfill #2}

\makeatletter%
\begin{document}

\title{When Lifetimes Liberate: A Type System for Arenas with Higher-Order Reachability Tracking}

\author{Siyuan He}
\orcid{0009-0002-7130-5592}
\affiliation{%
  \institution{Purdue University}
  \city{West Lafayette}
  \country{USA}
}
\email{he662@purdue.edu}

\author{Songlin Jia}
\orcid{0009-0008-2526-0438}
\affiliation{%
  \institution{Purdue University}
  \city{West Lafayette}
  \country{USA}
}
\email{jia137@purdue.edu}

\author{Yuyan Bao}
\orcid{0000-0002-3832-3134}
\affiliation{%
  \institution{Augusta University}
  \city{Augusta}
  \country{USA}
}
\email{yubao@augusta.edu}

\author{Tiark Rompf}
\orcid{0000-0002-2068-3238}
\affiliation{%
  \institution{Purdue University}
  \city{West Lafayette}
  \country{USA}
}
\email{tiark@purdue.edu}

\begin{abstract}

Statically enforcing safe resource management is challenging due to tensions between flexible lifetime disciplines and expressive sharing patterns.
Region-based systems offer lexically scoped regions under a stack discipline, wherein resources are managed in bulk.
In many such systems, however, resources are second-class and can neither escape their scope nor be freely returned from functions.
Ownership and linear type systems, such as Rust, offer first-class, non-lexical lifetimes with robust static guarantees, 
but rely on invariants that limit higher-order patterns and expressive sharing.

In this work, we propose a type system that uniformly treats all heap-allocated
resources under diverse lifetime, granularity, and sharing settings.
Our system provides programmers with three allocation modes: 
(1) fresh allocation for first-class, non-lexical resources;
(2) fresh allocation for second-class resources with lexically bounded lifetimes; and
(3) coallocation that groups resources by shadow arenas
for bulk tracking and deallocation.
Regardless of mode, resources are represented uniformly at the type level,
supporting generic abstraction and preserving the higher-order parametric
nature of the language.

Obtaining static safety in higher-order languages with flexible sharing is nontrivial. 
To address this, our solution builds on reachability types,
and our extension adds the capability to track both individual and grouped resources,
enables the expression of cyclic store structures,
and allows the selective enforcing of stack lifetime discipline.
These mechanisms are formalized in the \arenalang and \withlang type systems, 
which are proven type safe and memory safe in Rocq. \end{abstract}



\maketitle

\lstMakeShortInline[keywordstyle=,%
                    flexiblecolumns=false,%
                    language=PolyRT,
                    basewidth={0.56em, 0.52em},%
                    mathescape=true,%
                    basicstyle=\footnotesize\ttfamily]@

\newrulename{t1cst}{t-cst}
\newrulename{t1var}{t-var}
\newrulename{t1ref}{t-ref}
\newrulename{t1refat}{t-refat}
\newrulename{t1get}{t-deref}
\newrulename{t1put}{t-assgn}
\newrulename{t1sub}{t-sub}
\newrulename{t1loc}{t-loc}
\newrulename{t1abs}{t-abs}
\newrulename{t1app}{t-app}
\newrulename{t1appfr}{t-app$^{\starred{}}$}
\newrulename{t1tabs}{t-tabs}
\newrulename{t1tapp}{t-tapp}
\newrulename{t1tappfr}{t-tapp$^{\starred{}}$}

\newrulename{sqsub}{sq-sub}
\newrulename{strans}{s-trans}
\newrulename{srefl}{s-refl}
\newrulename{sref}{s-ref}
\newrulename{sfun}{s-fun}
\newrulename{stvar}{s-tvar}
\newrulename{sall}{s-all}
\newrulename{qsub}{q-sub}
\newrulename{stop}{s-top}
\newrulename{qcong}{q-cong}
\newrulename{qself}{q-self}
\newrulename{qvar}{q-var}
\newrulename{qtrans}{q-trans}
\newrulename{qqvar}{q-qvar}

\newrulename{t2cst}{t-cst}
\newrulename{t2var}{t-var}
\newrulename{t2ref}{t-ref}
\newrulename{t2refat}{t-refat}
\newrulename{t2get}{t-deref}
\newrulename{t2put}{t-assgn}
\newrulename{t2sub}{t-sub}
\newrulename{t2loc}{t-loc}
\newrulename{t1withr}{t-refin}
\newrulename{t2withr}{t-refin}
\newrulename{t2withc}{t-locin}
\newrulename{t2abs}{t-abs}
\newrulename{t2app}{t-app}
\newrulename{t2appfr}{t-app$^{\starred{}}$}
\newrulename{t2tabs}{t-tabs}
\newrulename{t2tapp}{t-tapp}
\newrulename{t2tappfr}{t-tapp$^{\starred{}}$}

\newrulename{eubase}{eu-base}
\newrulename{eufresh}{eu-fresh}
\newrulename{eukill}{eu-kill}

\newrulename{tscst}{ts-cst}
\newrulename{tsvar}{ts-var}
\newrulename{tsref}{ts-ref}
\newrulename{tsrefat}{ts-refat}
\newrulename{tsget}{ts-deref}
\newrulename{tsput}{ts-assgn}
\newrulename{tssub}{ts-sub}
\newrulename{tswithr}{ts-refin}
\newrulename{tsabs}{ts-abs}
\newrulename{tsapp}{ts-app}
\newrulename{tsappfr}{ts-app$^{\starred{}}$}
\newrulename{tstabs}{ts-tabs}
\newrulename{tstapp}{ts-tapp}
\newrulename{tstappfr}{ts-tapp$^{\starred{}}$}
\section{Introduction}

\looseness=-1
Higher-order functional languages, starting with {LISP} \cite{DBLP:conf/acm/000159}, typically rely on
garbage collection or reference counting for automatic resource management.
Although safe and convenient, these approaches offer little control over the timing or granularity of deallocation.
Such a trade-off is often acceptable for memory resources, but
is a key limitation for system assets like files, sockets, or large heaps.
Languages mitigate this trade-off with explicit APIs like @FileReader.close@
in Java or scoped constructs like @with@ blocks in Python.
Still, these patterns do not ensure safety \emph{statically}:
deallocated resources may be referenced directly
or leaked via global storage,
leading to runtime exceptions at best and
data corruption at worst in \emph{use-after-free} (UAF) situations.
An ideal solution should combine the convenience of automatic management 
with explicit control for critical resources, 
enforcing static safety like no UAF without restricting common higher-order expressiveness.

\paragraph{Stack and Heap Allocation.}
To recognize the challenge, recall the classic contrast between stack and heap allocation. 
Stack values have strictly lexical lifetimes and are reclaimed automatically when their scope ends,
guaranteeing safety and efficiency, but making them inherently \emph{second-class} where they cannot escape their defining scope, be returned, or be stored in a heap structure.
In contrast, heap allocation yields \emph{first-class} values that can flow freely through higher-order functions, either by being captured in closures or shared across structures.
However, such flexibility comes at the cost of timely deallocation by programmer control, while garbage collection or reference counting eventually reclaim objects, giving up the static safety of stack discipline. 

Works such as \citet{DBLP:conf/oopsla/OsvaldEWAR16}, which reintroduce second-class values in higher-order programs to provide memory safety and scoped lifetime control, have also been shown to be effective at reducing GC pressure \cite{xhebraj_et_al:LIPIcs.ECOOP.2022.15}. 
More recent efforts have revived interest in the area, for example, stoic functions \cite{stoic_functions} and a lambda calculus with second-class functions and references \cite{10.1145/3759427.3760373}.
We aim to combine the expressiveness of first-class heap objects with the static safety of stack allocation.

\paragraph{Regions with Scoped Lifetimes.}
Region-based systems \cite{MLkit_new,cyclone} can be viewed as an attempt to re-introduce these stack-like benefits within heap allocation. 
Programmers allocate groups of memory resources together into regions (arenas), and deallocate an entire region in bulk upon exiting a designated lexical scope, similar to @with@ blocks in Python.
Values can be shared freely within a region, but the region itself is lexically scoped.

\paragraph{Arenas: When Lifetime Liberate.}
Lexically scoped regions are not the only model for safe resource management, 
as individual, non-lexical lifetimes are also beneficial \cite{fluetImplementationPerformanceEvaluation},
particularly for patterns like event loops and garbage collectors.
\citet{DBLP:conf/iwmm/HicksMGJ04} further recharacterized regions as \emph{LIFO arenas},
highlighting that \emph{arenas}, dynamically growable collections for
bulk operation, can integrate diverse lifetime control strategies
to additionally achieve \emph{flexibility in lifetimes}.
Building on this insight, prior works \cite{first_class_region,verona}
formally explored arenas with linear types \cite{wadler1990linear}
and ownership types \cite{DBLP:conf/oopsla/ClarkePN98,DBLP:conf/ecoop/NobleVP98}.
These systems employ flow-sensitive reasoning, which necessitates a first-order setting where the global invariants effectively make all resources behave as second-class,
 restricting higher-order abstraction.
To maintain global invariants, users must explicitly \emph{open/enter} an arena
before accessing its content,
an operation similar to \emph{mutable borrows} in Rust \cite{rust} but at a higher level,
which ensures safe deallocation without enforcing LIFO lifetimes.

\begin{table}[tp]
\small
\caption{Comparison of resource tracking/management solutions.
Each solution emphasizes a distinct strategy, while
in this work, we try to unify their strengths on control, expressiveness, and flexibility.}\label{tbl:featcompare}
\begin{adjustbox}{width={\textwidth},keepaspectratio}
\begin{tabular}{lcccccc}
\toprule
                       & Cyclone$^\text{a}$ & Cyclone$^\text{b}$ & Safe Rust$^\text{c}$ & Reggio$^\text{d}$ & {$\weipolylang$}$^\text{e}$ & This Work \\
Features               & [LIFO] & [Dynamic] & [Ownership] & [Region+OT] & [Reachability] & [Region+RT] \\ \midrule
Stack Lifetime Discipline & \checkmark & \checkmark & \checkmark & \checkmark &            & \checkmark \\
Mutable, Cyclic Sharing & \checkmark  & \checkmark  &            & \checkmark &            & \checkmark \\
First-Class Escaping   &              &  \checkmark  & \checkmark & \checkmark & \checkmark & \checkmark \\
Higher-Order Functions & \checkmark &                & \checkmark &            & \checkmark  & \checkmark \\
\bottomrule\vspace{-1em} \\
\multicolumn{7}{l}{\footnotesize\,$^\text{a}$\citet{cyclone}\quad$^\text{b}$\citet{DBLP:conf/iwmm/HicksMGJ04}\quad$^\text{c}$\citet{rust}\quad$^\text{d}$\citet{verona}\quad$^\text{e}$\citet{wei24}}%
\end{tabular}
\end{adjustbox}
\vspace{-3ex}
\end{table}

\begin{figure}[t]
\begin{minipage}{.52\linewidth}  
\begin{lstlisting}
{
  val fr = new Ref(42)        //: Ref[Int]$\trackvar{fr}$
  val ar = new Ref(42) scoped //: Ref[Int]$\trackvar{ar}$
  val f1 = new Ref(42) at fr  //: Ref[Int]$\trackvar{f1,fr}$
  val a1 = new Ref(42) at ar  //: Ref[Int]$\trackvar{a1,ar}$
  val a2 = new Ref(42) at a1  //: Ref[Int]$\trackvar{a2,a1}$
}  // guaranteed to deallocate {ar,a1,a2}
\end{lstlisting}
\subcaption{An allocation can take one of three forms:
primitive (\lstinline|fr|), scoped (\lstinline|ar|),
and coallocation (\lstinline|f1,a1,a2|).
Scoped arenas (\eg, \lstinline|ar|) are deallocated when leaving the scope.}
\label{fig:intro-mm-snippet}
\end{minipage}
\hfill
\begin{minipage}{.45\linewidth}
\vspace{2ex}
\begin{tikzpicture}

  \tikzset{
    cellstyle/.style={draw, fill=orange!15, minimum width=1cm, minimum height=.6cm, anchor=center, text height=1.5ex, text depth=.25ex},
    dotstyle/.style={draw=none, fill=none, minimum width=1cm, minimum height=.6cm, anchor=center, text height=1.5ex, text depth=.25ex},
    boundstyle/.style={black, double=black, double distance=0.5pt, line width=0.3pt}
  }
  
  \matrix (a1) [matrix of nodes,
    row sep=0cm,
    nodes in empty cells,
    inner sep=0pt,
    anchor=north,
  ] at (0,0)
  {
    \node (a11) [cellstyle] {fr}; \\
    \node (a12) [cellstyle] {f1}; \\
    \node (a13) [dotstyle] {...}; \\
  };
  \draw[boundstyle]
      ([shift={(-0.0,-.6)}]a1.south west) --
      ([shift={(-0.0,0.0)}]a1.north west) --
      ([shift={(0.0,0.0)}]a1.north east) --
      ([shift={(0.0,-.6)}]a1.south east);

  \matrix (a2) [matrix of nodes,
    row sep=0cm,
    nodes in empty cells,
    inner sep=0pt,
    anchor=north,
  ] at ([xshift=1.8cm]a1.north)
  {
    \node (a21) [cellstyle] {ar}; \\
    \node (a22) [cellstyle] {a1}; \\
    \node (a23) [cellstyle] {a2}; \\
    \node (a25) [dotstyle] {...}; \\
  };
  \draw[boundstyle]
    ([shift={(-0.0,-0.0)}]a2.south west) --
    ([shift={(-0.0,0.0)}]a2.north west) --
    ([shift={(0.0,0.0)}]a2.north east) --
    ([shift={(0.0,-0.0)}]a2.south east);

  \matrix (a3) [matrix of nodes,
    row sep=0cm,
    nodes in empty cells,
    inner sep=0pt,
    anchor=north,
  ] at ([xshift=1.8cm]a2.north)
  {
    \node (a31) [dotstyle] {...}; \\
  };
  \draw[boundstyle, dashed]
    ([shift={(-0.0,-1.8)}]a3.south west) --
    ([shift={(-0.0,0.0)}]a3.north west) --
    ([shift={(0.0,0.0)}]a3.north east) --
    ([shift={(0.0,-1.8)}]a3.south east);

  \matrix (a4) [matrix of nodes,
    row sep=0cm,
    nodes in empty cells,
    inner sep=0pt,
    anchor=north,
  ] at ([xshift=1.4cm]a3.north)
  {
    \node (a41) [dotstyle] {...}; \\
  };

\end{tikzpicture}
 \vspace{1ex}
\subcaption{The two-dimensional layout produced by (a).
Two arenas are created; they can be referred to in \lstinline|at| clauses
through existing cells, \eg, \lstinline|fr,ar,a1|.}
\label{fig:intro-mm-layout}
\end{minipage}
\caption{Illustration of the three allocation forms and the corresponding
two-dimensional store layout.}
\vspace{-3ex}
\end{figure}

As summarized in \Cref{tbl:featcompare}, prior work spans diverse strategies for
resource tracking and management.
This work aims to unify their strengths:
the safety of stack discipline,
the expressiveness of higher-order languages,
and the treatment of resources as first-class entities.

\subsection{Flexible Resource Management in One Coherent System: Bulk and Individual Granularity, Lexical and Unbounded Lifetimes}

Our system treats heap-allocated mutable references as the canonical form of resources. 
These references are first-class values that may be passed, stored, or returned without restriction. 
Programmers retain control over resource lifetimes and granularity through three allocation forms:
\begin{enumerate}%
  \item allocation of a first-class, non-lexical reference;
  \item scoped allocation of a lexically bounded reference following stack discipline; and
  \item coallocation, which groups resources collectively into \emph{shadow arenas}.
\end{enumerate}

All three forms (\Cref{tbl:allocation-form}) coexist under a single uniform reference type @Ref[T]@, so the language does not distinguish resources following stack discipline from first-class resources at the type level. 
Built for a higher-order functional setting, our system supports shared mutable state and cyclic store structures without requiring flow-sensitive reasoning.

We illustrate our resource allocation forms in \Cref{fig:intro-mm-snippet}.
The @new@ allocations for @fr@ and @ar@ introduce two distinct references, and implicitly two distinct shadow arenas to host them.
Subsequent coallocations, @f1@, @a1@, and @a2@, coallocate additional resources in the same arena of placement clauses prefixed by @at@, avoiding the need for explicit handlers.
Scoped allocations, annotated with @scoped@, tie a reference and its shadow arena to the lexical scope, enabling automatic bulk deallocation of the entire shadow arena at scope exit.
Within their scopes, scoped and non-scoped resources behave uniformly, preserving the first-class abstraction of references while ensuring predictable reclamation where selectively desired.

\subsection{Type Safety and Safety of Deallocation in a Higher-Order
Setting: Mutable Sharing without Flow-Sensitive Reasoning}

In languages with first-class references, higher-order functions, and mutable sharing, statically preventing errors like UAF is challenging.
Our system enforces static safety through two complementary mechanisms, corresponding to non-lexical and lexical lifetime control.

\paragraph{First-class safety via reachability tracking.}
We build on \emph{reachability types} \cite{bao21,wei24,david} to track first-class resources that persist non-lexically.
Each type carries a qualifier signifying its reachable resources, marking sharing and separation. 
We propose \arenalang that extends prior reachability system $\weipolylang$ \cite{wei24} with shadow arenas, reference coallocation (@new at@ in \Cref{fig:intro-mm-snippet}), and a two-dimensional store model (visualized in \Cref{fig:intro-mm-layout}).
\arenalang enables collective tracking and reasoning of resources within an arena.
By grouping resources within an arena under a coarse-grained reachability relation, \arenalang achieves greater expressiveness, supporting more complicated store topologies (see \Cref{sec:store-topo}), 
while preserving the static safety guarantees as earlier fine-grained approaches.

\paragraph{Selective stack discipline for scoped resources.}
Programmers may also declare an arena as scoped (@new scoped@ in \Cref{fig:intro-mm-snippet}), 
causing it to follow a stack discipline to ensure automatic deallocation at scope exit.
Building atop \arenalang, \withlang introduces such scoped allocation and statically guarantees that there is no UAF error at runtime, while maintaining flow-insensitive reasoning.
The scoping control is orthogonal to arenas and generalizable to non-memory resources, yet does not impose any type distinction between first-class and scoped resources.

\subsection{Contributions}

We propose a unified and flexible system for resource management in higher-order languages with mutable sharing, which combines the strengths of different flavors of resource management.
Specifically, our system is novel in the combination of:
\begin{enumerate}[leftmargin=1.6em]
  \item Unified treatment of resources: All memory resources are \textbf{first-class} values with a \textbf{single reference type} that abstracts over stack-like and heap allocation. This allows client code to be generic over the particular storage model of references.
  \item Flexibility in control: Users can manage memory resources lexically or non-lexically, individually or collectively, without introducing type distinctions.
  \item Static safety guarantees: Reachability types track the flow of memory resources and guarantee their safe use. Users can impose selective \textbf{stack discipline} to guarantee predictable deallocation and no use-after-free errors through \textbf{flow-insensitive} reasoning.
  \item Expressive higher-order features: Our system supports higher-order functions with mutable sharing and cyclic store structures, surpassing the expressiveness of prior reachability systems.
\end{enumerate}
To the best of our knowledge, our system is the first to implement the intersection of these properties in a unified system, without separating first-class and second-class memory resources.
It is also the first work to provide a formal deallocation reasoning in this setting, supporting cyclic store structures while preserving static safety guarantees among similar systems.

In this work, we propose a new solution for resource management in higher-order
languages with mutable sharing, by integrating \emph{shadow} arenas and scoped lifetime reasoning atop reachability types.
Our system offers added flexibility in store topologies compared with previous reachability types, introducing additional scoped or coallocated introduction forms of store objects without type distinction between forms.
Specifically,
\begin{itemize}[leftmargin=1.6em]
  \item After briefly reviewing reachability types, we informally discuss the telescoping structures and the store topology in our system with added flexibility (\Cref{sec:overview-semantic}).
  \item We informally overview the $\arenalang$-calculus and the $\withlang$-calculus, discussing shadow arenas, coarse-grained reachability tracking, and relaxation of telescoping structures (\Cref{sec:overview}).
  \item We present the formal theory and metatheory of the \arenalang-calculus, a reachability type system with non-lexical arenas, coarse-grained reachability tracking, and a two-dimensional store semantics (\Cref{sec:formal}).
  \item We propose the \withlang-calculus atop \arenalang-calculus with scoped arena allocation and flow-insensitive deallocation reasoning, and present its formal theory and metatheory (\Cref{sec:scope-formal}).
  \item We present three case studies, on general fixed point combinators (\Cref{sec:case-fixpoint}), callback registration via non-lexical arenas (\Cref{sec:case-callback}), and cyclic store structures (\Cref{sec:case-multihop}).
\end{itemize}

In \Cref{sec:discussion} we reflect on limitations, design choices, and directions for future and parallel work.
\Cref{sec:related-word} surveys related work, and \Cref{sec:conclusion} wraps up the paper.
All formal results have been mechanized in Rocq, available online at \url{https://github.com/tiarkrompf/reachability} .

\section{Store Topology and Safety Through Reachability} \label{sec:overview-semantic}

Before introducing the technical details of our system, we briefly introduce reachability types and discuss how we achieve safety guarantees through reachability tracking.
Then we discuss the store topology and store invariants in our system, including a comparison to Rust and previous reachability type systems.

\subsection{A Review of Reachability Types} \label{sec:overview-RTrecap}

Reachability types \cite{bao21,wei24,david} aim to provide static safety guarantees in higher-order functional programs by tracking aliasing and its absence: separation.

\subsubsection{Tracking Resources.} 
The key idea of reachability types is to track which resources a value may reach.
A type\footnote{For bindings in snippets, \lstinline|val x = t // : T $\phantom{n}\dashv$ $\phantom{n}\Gamma$|, we illustrate \lstinline|$\Gamma\phantom{.}$$\vdash\phantom{.}$ x : T| unless noted otherwise. \lstinline|$\Gamma$| might be omitted. } is annotated with a reachability \emph{qualifier}\footnote{We omit qualifiers for untracked values, including integers and booleans.} recording the static \emph{variables} the term may reach.
We mark a reference as \emph{fresh} with a marker $\QFresh$ in its qualifier if the reference has not yet been bound to a variable.
A reference allocation @new Ref(7)@ bound to variable @r@ is typed as,
\begin{lstlisting}
  val r = new Ref(7)          // new Ref(7) : Ref[Int]$\trackfresh$
  r                           // r : Ref[Int]$\trackvar{r}$
\end{lstlisting}
At the runtime, the @new Ref@ expression will evaluate to a fresh \emph{location} in the store, and qualifiers additionally track \emph{locations} in dynamic typing.
\begin{lstlisting}
  // new Ref(7) evaluates to some $\ell$ in the runtime
  $\ell$                            // $\ell$ : Ref[Int]$\trackvarm{\ell}$  where $\sigma(\ell)$ = 7  in store $\sigma$
\end{lstlisting}

Since the runtime store location of a reference is unknown in the static program,
reachability types conservatively approximate runtime locations in static typing.
Variables and locations are not necessarily in one-to-one correspondence.
For instance, in a conditional statement that reaches references in both branches,
\begin{lstlisting}
  val rx = new Ref(42)        // : Ref[Int]$\trackvar{rx}$
  val ry = new Ref(7)         // : Ref[Int]$\trackvar{ry}$
  if (cond) rx else ry        // : Ref[Int]$\trackvar{rx, ry}$
\end{lstlisting}
The static qualifier of the conditional must conservatively include both @rx, ry@, even though exactly one location will be produced at runtime.
In general, a variable in a reachability qualifier represents an \emph{over-approximation} of the set of locations that are actually reached.

\subsubsection{Aliases and Closures.}
Multiple variables can bind to the same value from aliasing. %
\begin{lstlisting}
  val r = new Ref(7)          // : Ref[Int]$\trackvar{r}$ $\dashv$ [ r : Ref[Int]$^\qfresh$ ]
  val s = r                   // : Ref[Int]$\trackvar{s}$ <: Ref[Int]$\trackvar{r}$ $\dashv$ [ r : Ref[Int]$^\qfresh$, s : Ref[Int]$\trackvar{r}$ ]
\end{lstlisting}
Reachability types initially assign each variable a minimal qualifier containing only itself.
Through subtyping (\Cref{sec:formal-typing-subtyping}), the qualifier of variable @s@ can upcast to its recorded qualifier in the typing context, which is @r@, the variable that @s@ aliases.

Higher-order functions introduce closures that may capture resources.
A function qualifier specifies the resources required for the computation. 
\begin{lstlisting}
  def f1() = !r                 // : (() => Int)$\trackvar{f1}$ $\dashv$ [ f1 : (() => Int)$\trackvar{r}$ ]
  val f2 = {
    val x = new Ref(6)          // : Ref[Int]$\trackvar{x}$
    () => x                     // : (() => Ref[Int]$\trackvar{x}$)$\trackvar{x}$
  } // : (() => Ref[Int]$\trackvar{f2}$)$\trackvar{f2}$
\end{lstlisting}
The function qualifier of @f1@ in the typing context records the free variables captured by @f1@.
Closure @f2@ returns a nameless function capturing a locally allocated reference @x@, which loses its name at scope exit and degrades to the self-reference of the closest closure @f2@. 
Reachability types preserve tracking of resources \emph{escaping} their lexical scopes via function self-references.

\subsubsection{Separation and Controlled Sharing.}
With reachability tracked, functions can require the separation of resources between the argument and the function by the domain qualifier.
A freshness marker $\QFresh$ in the function domain qualifier requires that the argument is contextually fresh relative to the function, 
which ensures separation between computations.
Controlled sharing can be specified through a function domain qualifier @$\QFresh$q@, permitting at most the overlap @q@ observable from both the function and the argument.
\begin{lstlisting}
  val u = new Ref(42)               // : Ref[Int]$\trackvar{u}$
  val v = new Ref(41)               // : Ref[Int]$\trackvar{v}$
  def f(x : Ref[Int]$\trackvar{\qfresh u}$) = !u + !v     // : (Ref[Int]$\trackvar{\qfresh u}$ => Int)$\trackvar{u,v}$
  f(u)                              // OK, {u} $\cap$ {u,v} $\subseteq$ {$\qfresh$u}
  f(v)                              // $\color{red}{\texttt{Error:}}$ {v} $\cap$ {u,v} 
\end{lstlisting}
The domain qualifier @$\QFresh$u@ of function @f@ permits the argument to share only @u@ with the function, but not @v@, achieving the sense of controlled sharing.

\subsection{Telescoping Structures} \label{sec:overview-telescope}
As a common issue in dependent type systems, \emph{telescoping structures} arise in reachability types where earlier allocated store objects cannot reach later ones.
Such asymmetrical store structures prevent previous reachability type systems from representing cyclic store structures, including mutual recursion.

Telescoping structures are reflected in the static typing, as the order of declared variables approximates the allocation timeline.
Consider extending \emph{Landin's Knots} to two mutually recursive references capturing each other, 
\begin{lstlisting}
  val c1 = ...                     // : Ref[(Int => Int)$\trackvar{q1}$]$\trackvar{c1}$
  val c2 = ...                     // : Ref[(Int => Int)$\trackvar{q2}$]$\trackvar{c2}$
  def f1(x : Int) = (!c1)(x)       // : (Int => Int)$\trackvar{c1}$, captures c1
  def f2(x : Int) = (!c2)(x)       // : (Int => Int)$\trackvar{c2}$, captures c2
  c2 := f1                         // OK, q2 can observe c1
  c1 := f2                         // $\color{red}{\texttt{Error:}}$ q1 cannot observe c2
\end{lstlisting} 
The first assignment @c2 := f1@ is type-checked because @q2@ can include @c1@, which is in the context at the declaration of @c2@.
The second assignment @c1 := f2@ triggers a type error because @q1@ cannot include @c2@, as @q1@ is determined before @c2@ is declared.

Telescoping structures forbid all store cycles in previous reachability types, preventing the encoding of complex data structures such as cyclic linked lists. 
In our work, the issue is relaxed through \emph{coarse-grained reachability tracking} that reasons about separation and sharing at the arena level.
This permits cyclic store structures both within and across arenas, as further illustrated in the store topology (\Cref{sec:store-topo}) and static typing (\Cref{sec:overview-relax-telescope}).

\subsection{Store Topology}

Languages with static safety guarantees typically impose invariants on the shape of the store to ensure tractable reasoning. 
These invariants define which forms of sharing, mutation, and lifetime behaviors are possible in permitted programs. 
Comparing the topological restrictions across systems sheds light on the design trade-offs underlying their safety models.

We begin by reviewing Rust as a well-known instance of ownership types, then discuss the prior work on reachability types, and finally introduce the store topology of our work.

\begin{figure}
    \begin{mdframed}
    \begin{subfigure}[t]{.48\textwidth}
        \centering
        \begin{tikzpicture}[node distance = 1.2cm]

            \node[locs] (N1) at (current page.east) {};
                \node[above=.05cm of N1] {\texttt{Root}};

            \node[locs, above left of = N1, yshift=-0.25cm] (N2) {};
            \node[locs, above left of = N2, yshift=-0.25cm, xshift=-0.5cm] (N3) {};
            \node[locs, below left of = N2, yshift=0.5cm, xshift=-0.25cm] (N4) {};

            \node[locs, below left of = N1, xshift=-0.5cm, yshift=0.25cm] (N5) {};
            \node[locs, below left of = N5, xshift=-0.5cm, yshift=0.25cm] (N6) {};

            \node[locs] (N0) at ([xshift=0.5cm]N1.south east|-N6.center) {};

            \draw[nodearrow] (N1) -- (N2);
            \draw[nodearrow] (N2) -- (N3);
            \draw[nodearrow] (N2) -- (N4);
            \draw[nodearrow] (N1) -- (N0);
            \draw[nodearrow] (N1) -- (N5);
            \draw[nodearrow] (N5) -- (N6);
            \draw[-latex, red] (N5) -- (N4)
                node[draw, midway, -, red, sloped, cross out, line width=.5ex, minimum width=1.5ex, minimum height=1ex, anchor=center]{};

            \draw[->, thick] ([xshift=-0.5cm, yshift=-0.5cm]N6.west|-N0.south east) -- ([xshift=0.5cm, yshift = -0.5cm]N0.south east) node[midway, below] {Time};
        \end{tikzpicture}
        \caption{The store topology in Rust is a tree structure without \emph{telescoping} over time. The directed edges represent \emph{ownership} of the inbound location where each location is uniquely owned by its parent. Later allocated locations can be owned by earlier allocated locations, but shared ownership is not permitted. }
        \label{fig:intro-topo-rust}
    \end{subfigure}
    \hfill
    \vline
    \hfill
    \begin{subfigure}[t]{.48\textwidth}
        \centering
        \begin{tikzpicture}[node distance = 1.2cm]
            \node[locs] (N1) at (current page.east) {};

            \node[locs, above left of = N1, yshift=-0.25cm] (N2) {};
            \node[locs, above left of = N2, yshift=-0.25cm, xshift=-0.5cm] (N3) {};
                \node[right=.05cm of N3, yshift=0.25cm] {\small self-loop};
            \node[locs, below left of = N2, yshift=0.5cm, xshift=-0.25cm] (N4) {};
                \node[left=.05cm of N4] {\small shared};

            \node[locs, below left of = N1, xshift=-0.5cm, yshift=0.25cm] (N5) {};
            \node[locs, below left of = N5, xshift=-0.5cm, yshift=0.25cm] (N6) {};

            \node[locs] (N0) at ([xshift=0.5cm]N1.south east|-N6.center) {};

            \draw[nodearrow] (N1) -- (N2);
            \draw[nodearrow] (N2) -- (N3);
            \draw[nodearrow] (N2) -- (N4);
            \draw[nodearrow] (N1) -- (N5);
            \draw[nodearrow] (N5) -- (N6);
            \draw[nodearrow] (N5) -- (N4);
            \draw[loop] (N3) to[out=135,in=-135] (N3);
            \draw[-latex, red] (N1) -- (N0)
                node[draw, midway, -, red, sloped, cross out, line width=.5ex, minimum width=1.5ex, minimum height=1ex, anchor=center]{};

            \draw[->, thick] ([xshift=-0.5cm, yshift=-0.5cm]N6.west|-N0.south east) -- ([xshift=0.5cm, yshift = -0.5cm]N0.south east) node[midway, below] {Time};
        \end{tikzpicture}
        \caption{The store topology of reachability types is a directed acyclic graph with self-loops. The directed edges represent reachability from the outbound location to the inbound one. Shared reachability of a location is allowed. \emph{Telescoping} ensures that no location can reach later allocated locations.  }
        \label{fig:intro-topo-rt}
    \end{subfigure}
    \caption{The comparison of store topology between Rust \cite{rust} (left) and reachability type \cite{david} (right). Circular nodes in purple represent store locations. Directed edges represent the references from one location to another, ruled by either ownership or reachability, respectively. The timeline indicates the order of allocation in the program flow. }
    \label{fig:intro-topo}
    \end{mdframed}
    \vspace{-2ex}
\end{figure} 
\subsubsection{Store Topology in Rust.}

Rust's memory model enforces strict, statically checked ownership, where each heap-allocated object has exactly one unique owner. 
These ownership relationships induce a tree-shaped object graph (\Cref{fig:intro-topo-rust}), where each object has at most one inbound edge from its owner.
In the store topology, each ownership edge grants exclusive control to read, write, and deallocate the corresponding object.

Rust also supports borrowed references (B@&@ or @&mut@) that grant temporary access without transferring ownership.
Borrows neither extend the owner's lifetime nor alter the ownership hierarchy.
The borrow checker statically ensures that all borrows are strictly bounded by the lifetime of the original owner, and cannot outlive the owner to cause use-after-free or dangling reference errors.

Deallocation of an object is triggered automatically when the unique owner goes out of scope, with all associated borrows guaranteed to have ended beforehand.
This tree-shaped discipline yields predictable, statically enforced deallocation, but restricts flexible sharing and cyclic structures.

\subsubsection{Store Topology in Reachability Types.} \label{sec:overview-RT-topo}

The store topology in reachability types is more permissive than Rust's, as objects may have multiple inbound edges enabling shared access, as shown in \Cref{fig:intro-topo-rt}.
However, reachability types have their own limitation that telescoping structures (\Cref{sec:overview-telescope}) prevent cyclic store structures.

While reachability types enable static reasoning about separation and aliasing, they do not provide an integrated treatment of deallocation. 
Existing systems assume that unreachable resources are eventually reclaimed by garbage collection.
As potential extensions, prior work \cite{bao21,wei24} has informally proposed, but not implemented, flow-sensitive effect systems to reason about safe deallocation.

\begin{figure}
  \begin{mdframed}
  \begin{subfigure}{.48\textwidth}
    \centering

    \begin{tikzpicture}[node distance = 1.2cm]

      \node[locs] (N1) at (current page.east) {};
          \node at (N1) {\small $c$};

      \node[locs, above left of = N1, yshift=0.15cm, xshift=0.65cm] (N2) {};
          \node at (N2) {\small $b$};

      \node[locs, left of = N1, xshift=0.5cm] (N32) {};
          \node at (N32) {\small $a$};

      \node[locs, below left of = N2, yshift=0.5cm, xshift=-0.5cm] (N41) {};
      \node[locs, above left of = N41, yshift=0cm, xshift=0.5cm] (N42) {};
      \node[locs, below left of = N41, yshift=0.5cm, xshift=-0.15cm] (N43) {};
      \node[locs, above left of = N41, yshift=-0.25cm, xshift=-0.5cm] (N44) {};

      \node[locs, below left of = N43] (N5) {};

      \node[locs, below left of = N32, xshift=-0.65cm, yshift=-0.25cm] (N01) {};
      \node[locs] (N02) at ([xshift=-1.5cm]N1.south east|-N01.center) {};

      \begin{scope}[on behind layer]
        \node[arena, fit=(N1) (N32)] (A1) {};
        \node[arena, fit=(N2)] (A2) {};
        \node[arena, fit=(N41) (N42) (N43) (N44)] (A4) {};
        \node[arena, fit=(N01) (N02)] (A0) {};
        \node[arena, fit=(N5)] (A5) {};
      \end{scope}

      \draw[nodearrow, looseness=0] (N1) to[out=70, in=-55] (N2);
      \draw[nodearrow, looseness=0] (N2) to[out=-90, in=102] (N1);
      \draw[nodearrow] (N32) -- (N01);
      \draw[nodearrow] (N2) -- (N32);
      \draw[nodearrow] (N43) -- (N5);
      \draw[nodearrow] (N44) -- (N5);
      \draw[nodearrow] (N32) -- (N41);
      \draw[nodearrow] (N32) -- (N43);

      \draw[nodearrow, teal, looseness=0] (N41) to[out=90, in=-55] (N42);
      \draw[nodearrow, teal, looseness=0] (N42) to[out=-90, in=125] (N41);
      \draw[nodearrow, teal, looseness=0] (N41) to[out=195, in=35] (N43);
      \draw[nodearrow, teal, looseness=0] (N43) to[out=0, in=-130] (N41);
      \draw[nodearrow, teal, looseness=0] (N42) to[out=-135, in=95] (N43);
      \draw[nodearrow, teal, looseness=0] (N43) to[out=60, in=-100] (N42); 
      \draw[nodearrow, teal] (N41) -- (N44);
      \draw[nodearrow, teal] (N42) -- (N44);
      \draw[nodearrow, teal] (N43) -- (N44);
      \draw[loop, teal] (N5) to[out=0,in=-90] (N5);

      \draw[nodearrow, teal, looseness=0] (N02) to[out=170, in=10] (N01);
      \draw[nodearrow, teal, looseness=0] (N01) to[out=-20, in=194] (N02);

      \draw[->, thick] ([xshift=-0.25cm, yshift=-0.5cm]N5.west|-N01.south east) -- ([xshift=2.5cm, yshift = -0.5cm]N01.south east) node[midway, below] {Time};
  \end{tikzpicture}

  \caption{ %
    The store topology of this work. Internal reachability ({\color{teal} teal}) is unrestricted within arenas and allows arbitrary structures. Cross-arena reachability (black) can be forward in time to subsequently coallocated locations, supporting cross-arena cycles. 
  }
  \label{fig:intro-topo-this}
\end{subfigure}
\hfill
\vline
\hfill
\begin{subfigure}{.48\textwidth}

\begin{lstlisting}[xleftmargin=.07\textwidth,numbers=left]
val a = new Ref(...)     
      // : Top$\trackvar{a}$
val b = new Ref(a)       
      // : Ref[Top$\trackvar{a}$]$\trackvar{b}$
val c = new Ref(b) at a  
      // : Ref[Top$\trackvar{b}$]$\trackvar{c}$ <: Top$\trackvar{a}$
b := c
\end{lstlisting}

  \caption{The code illustrates a cross-arena cycle between nodes \texttt{b} and \texttt{c} (left). Type details are eliminated via a top type. Reference \texttt{b} to \texttt{a} collectively reaches all locations in \texttt{a}, which allows \texttt{c}, the reference to \texttt{b}, to be assigned to \texttt{b}, forming a cross-arena cycle. }
  \label{fig:intro-topo-crosscycle}

\end{subfigure}
\caption{The store topology of this work (left) and illustration of cross-arena cycles (right). Arenas are marked as boxes containing multiple store locations filled with light {\color{orange!75} orange}. Internal reachability is marked in {\color{teal} teal}. Telescoping structures and exclusivity are relaxed in our work.  }
\label{fig:intro-topo-thiswork}
\end{mdframed}
\vspace{-2ex}
\end{figure} 
\subsubsection{Topology in Our Work.}   \label{sec:store-topo}

As shown in \Cref{fig:intro-topo-this}, our system generalizes the topologies by supporting cross-arena sharing and allowing arbitrary internal structures within each arena.
Unlike earlier reachability systems that track dependencies at the granularity of individual references, we employ a \emph{coarse-grained} (\Cref{sec:overview-coarse-grained}) model that tracks reachability collectively at the arena level.

All objects in the same arena share a common reachability identity, which eliminates telescoping constraints and enables arbitrary intra-arena object graphs and cyclic structures across arenas (\Cref{fig:intro-topo-crosscycle}).
This relaxation preserves reasoning about separation and safety while substantially expanding expressiveness.

Our system further introduces sound, bulk, and guaranteed deallocation through scoped allocation. 
Unlike Rust, which statically enforces that no reference outlives its owner, 
our system permits references to deallocated objects to persist.
Inbound edges to deallocated arenas are not cut off in the store topology, permitting references to retain reachability to objects that are no longer alive.
Static typing ensures that dereferencing these references cannot access deallocated resources, thereby maintaining memory safety under flexible lifetime control.
\section{Informal Overview} \label{sec:overview}

This section informally introduces the \arenalang-calculus, which features non-lexical arenas (\Cref{sec:overview-arenas}), and the \withlang-calculus, which extends \arenalang with scoped resource management and deallocation reasoning (\Cref{sec:overview-scopes}).

\subsection{Arenas: Shadow and Non-Lexical} \label{sec:overview-arenas}

Starting from \arenalang, we focus on non-lexical shadow arenas and coarse-grained reachability tracking.

\subsubsection{Shadow Arenas.}
We introduce the notion of \emph{shadow arenas}, describing that arenas in our system have no explicit names or constructors.
Instead, they are implicitly identified through references in the surface syntax.
Arenas are thus not real citizens in our language, but rather as abstract collections of references.
A dummy reference can serve as a sufficient proxy to be passed around explicitly when only the arena is required, such as for coallocation.

A new reference allocation (@new Ref@) implicitly establishes a fresh shadow arena containing only that reference.
Subsequent coallocations of form (@new Ref(..) at r@) place new references in the same shadow arena as the one of the \emph{proxy} reference @r@, even though the shadow arena itself is never explicitly named.

\begin{table}[h]
\small
\caption{Comparison of different reference allocation forms in our system, and how they interact with arenas and lifetime control. Only a qualifier but no type distinction exists among the three introduction forms. }\label{tbl:allocation-form}
\begin{tabular}{lllcccc}
\toprule
Form               & Syntax & Formal & Ref Type & Arena & Lifetime & \\ \midrule
(fresh) allocation & @new Ref(t)@ & $\tref~t$ & @Ref[T]$\trackfresh$@ & fresh &  non-lexical  \\
coallocation      & @new Ref(t) at r@ & $\trefat{t}{r}$  & @Ref[T]$\trackvar{r}$@ & same as @r@ & same as @r@   \\
scoped allocation & @new Ref(t) scoped@ &  $\twithr[\_]{t}{b}$  & @Ref[T]$\trackfresh$@ & fresh & lexical in $b$ \\
\bottomrule\vspace{-1em} \\
\end{tabular}
\vspace{-1.5ex}
\end{table}

As summarized in \Cref{tbl:allocation-form}, multiple allocation forms interact differently with the shadow arenas, yet the references allocated have a unified reference type. 
The different outermost reachability qualifiers implicitly encode the behavioral differences between fresh allocation and coallocation.
\begin{lstlisting}
  val a0 = new Ref(7)             // : Ref[Int]$\trackvar{a0}$ $\dashv$ [ a0 : Ref[Int]$\trackfresh$ ]
  // fresh arena containing {a0}
  val a1 = new Ref(8) at a0       // : Ref[Int]$\trackvar{a1}$ $\dashv$ [ a1 : Ref[Int]$\trackvar{a0}$, a0 : Ref[Int]$\trackfresh$ ]   
  // arena containing {a1,a0}
  val a2 = new Ref(9) at a1       // : Ref[Int]$\trackvar{a2}$ <: Ref[Int]$\trackvar{a1}$ <: Ref[Int]$\trackvar{a0}$
  // arena containing {a2,a1,a0}
\end{lstlisting}
The fresh allocation of @a0@ declares a fresh arena of a single reference cell, and thus the reference @a0@ is marked fresh in the typing context.
Coallocating @a1@ @at@ @a0@ yields a reference that inherits the reachability qualifier of its proxy reference @a0@, since both of them will reside in the same shadow arena. 
Hence, @a0@ and @a1@ are equivalent identifiers for that arena; coallocating @a2@ @at@ either reduces to the same location at the runtime.

\subsection{Coarse-grained Reachability Tracking} \label{sec:overview-coarse-grained}

\subsubsection{Two-Dimensional Stores and Bulk Reasoning.}
Recap that in the previous reachability systems, a dynamic store location $\ell$ evaluated from a reference can be qualified with $\ell$ itself, yielding \emph{location individuality} that qualifiers of all store locations are strictly disjoint from each other.
Thus, the one-dimensional store model is unsuitable for collective storage or bulk reasoning.

We generalize the store model to two-dimensional, indexing each reference by both its arena \emph{location} and an intra-arena \emph{offset}, denoted conventionally as $\lot{\ell}{o}$.
Reachability is tracked coarsely at the arena level (through the location $\ell$).

\begin{lstlisting}
  val a0 = new Ref(7)             // reduce to $\lot{\ell}{o_1}$ : Ref[Int]$\trackvarm{\ell}$ for some fresh $\ell$, $o_1$
  val a1 = new Ref(8) at a0       // reduce to $\lot{\ell}{o_2}$ : Ref[Int]$\trackvarm{\ell}$ for some fresh $o_2$
  val a2 = new Ref(9) at a1       // reduce to $\lot{\ell}{o_3}$ : Ref[Int]$\trackvarm{\ell}$ for some fresh $o_3$
\end{lstlisting}
Although the indexes of reduced reference cells differ in their offsets, they share the same reachability $\ell$, as they are collectively located in the same arena, allowing bulk reasoning.

\subsubsection{Connecting Static and Dynamic Typing.}
The static typing rule, which assigns a coallocated reference the same qualifier as its proxy reference (prefixed by @at@), reflects the coarse-grained view.
Consider a partially reduced program segment,
\begin{lstlisting}
  // after reducing a0 to $\lot{\ell}{o_1}$ : Ref[Int]$\trackvarm{\ell}$
  val a1 = new Ref(8) at ${\color{black}\lot{\ell}{o_1}}$         // : Ref[Int]$\trackvar{a1}$ $\dashv$ [ a1 : Ref[Int]$\trackvarm{\ell}$ ]
  a1                              // : Ref[Int]$\trackvarm{\ell}$  through subtyping 
\end{lstlisting}
Statically, coallocated references are treated as aliases of their proxy references, since they reduce to cells in the same arena, while static typing of a reference approximates the entire arena coarsely.
Such dependency in reachability yields two key features of shadow arenas: 
(1) any reference can serve as a proxy for its arena; and 
(2) reasoning about a reference implicitly applies to all objects within its arena.

\subsubsection{Escaping of Non-lexical Arenas.} \label{sec:overview-nonlexical-escaping}
A notable advantage of shadow arenas is to disentangle the reasoning about \emph{escaping} references from reasoning about their arenas.
References can safely outlive the lexical scope in which they are allocated, without requiring arenas to be passed or named explicitly.
When captured by closures, both references and arenas become \emph{non-lexical}, yet the type system should preserve their tracking for safety.

\begin{figure}[H]
\vspace{-2ex}
\begin{minipage}[t]{1\textwidth}
\begin{subfigure}[t]{.48\textwidth}
\begin{lstlisting}[xleftmargin=.05\textwidth,numbers=left]
def c1() = {
  val a = new Ref(())
  new Ref(7) at a
} // : (Unit => Ref[Int]$\trackfresh$)
c1() // : Ref[Int]$\trackfresh$
\end{lstlisting}
\end{subfigure}
\hfill
\vline
\hfill
\begin{subfigure}[t]{.48\textwidth}
\begin{lstlisting}[xleftmargin=.07\textwidth,numbers=left]
val c2 = { 
  val b = new Ref(())
  () => { new Ref(8) at b }
} // : (Unit => Ref[Int]$\trackvar{c2}$)
c2() // : Ref[Int]$\trackvar{c2}$
\end{lstlisting}
\end{subfigure}
\end{minipage}
\caption{Comparison between escaping of a reference in a fresh arena (left) and an existing arena (right).}
\label{fig:escaping-examples}
\vspace{-2ex}
\end{figure}

To illustrate the challenges of typing non-lexical arenas, consider the example of two closures wrapping different coallocation patterns as shown in \Cref{fig:escaping-examples}.
Each invocation of @c1@ allocates a fresh arena, while the closure @c2@ captures an existing arena.
The type system must distinguish the two patterns. 
In our design, the reachability qualifier of a coallocated reference inherits that of its proxy (reflecting the entire arena); 
reasoning about escaping coallocated references reduces to the standard reachability reasoning about escaping references (\Cref{sec:overview-RTrecap}).

\subsection{Relaxation of Telescoping Structures} \label{sec:overview-relax-telescope}
Coarse-grained reachability tracking also relaxes the telescoping structures in prior reachability systems, permitting multi-hop cycles through the store.
The coarse-grained reachability allows tracking future resources not yet coallocated in an arena without enlarging the reachability qualifier, as the entire arena shares the same reachability.

We illustrate the relaxation by showcasing the mutually recursive example, which is not allowed in previous reachability type systems, as discussed in \Cref{sec:overview-telescope}.
\begin{lstlisting}
  val a = new Ref(())             // : Ref[Unit]$\trackvar{a}$
  val c1 = ... at a               // : Ref[(Int => Int)$\trackvar{a}$]$\trackvar{c1}$ <: Ref[(Int => Int)$\trackvar{a}$]$\trackvar{a}$
  val c2 = ... at a               // : Ref[(Int => Int)$\trackvar{a}$]$\trackvar{c2}$ <: Ref[(Int => Int)$\trackvar{a}$]$\trackvar{a}$
  c1 := (x : Int) => (!c1)(x)     // OK. RHS : (Int => Int)$\trackvar{c1}$ <: (Int => Int)$\trackvar{a}$
  c2 := (x : Int) => (!c2)(x)     // OK.
\end{lstlisting} 
By coallocating @c1@ and @c2@ in the same arena @a@, they can reach each other via the reachability to the entire arena. 
More generally, arenas support arbitrary internal store structures, including self-loops and cyclic store structures.
Our system supports sound reasoning about such cyclic structures as a whole, such as separation and safe deallocation (\Cref{sec:case-multihop}). 

Using the same function capturing techniques, cross-arena multi-hop cycles can also be encoded, as illustrated in \Cref{fig:intro-topo-crosscycle}.

\subsection{Scopes: Bulk and Guaranteed Deallocation} \label{sec:overview-scopes}
While reachability types reason about aliasing and separation, they provide no mechanism for user-controlled, timely resource reclamation.
The \withlang-calculus extends \arenalang with scoped resource management, supporting automatic and bulk deallocation tied to lexical scopes.
Importantly, these deallocation guarantees are lightweight: they require no explicit effect tracking and are enforced solely upon the surface-level @scoped@ annotation.

\subsubsection{Guaranteed Deallocation.}
To regain user control of lifetime, we extend the \arenalang-calculus with \emph{scoped} allocation, 
whose lifetimes are bounded by their lexical blocks.
\begin{lstlisting}
  {
    val a = new Ref(()) scoped    // : Ref[Unit]$\trackvar{a}$
    !a                            // : Unit
  } // {a} is deallocated
\end{lstlisting}
The @scoped@ annotation marks the reference @a@ and subsequent references coallocated @at@ @a@ for automatic deallocation at the end of the block.
Specifically, as summarized in \Cref{tbl:allocation-form}, scoped allocation produces references having no type distinction from those created by fresh allocation, 
behaving no differently within their scope.
The type system ensures that such references cannot escape their defined scopes, preserving memory safety and predictable reclamation.

\subsubsection{Bulk Deallocation.}
When a scoped reference is deallocated, all resources coallocated within its arena are released together.
\begin{lstlisting}
  val sc2 = {
    val a = new Ref(()) scoped    // : Ref[Unit]$\trackvar{a}$
    val u = new Ref(42) at a      // : Ref[Int]$\trackvar{u}$
    val v = new Ref(!u)           // : Ref[Int]$\trackvar{v}$
    new Ref(v)                    // : Ref[Ref[Int]$\trackvar{v}$]$\trackfresh$
  } // {a,u} are deallocated
  !sc2                            // OK
\end{lstlisting}
To ensure static safety, the type system prohibits scoped references in either the type or the qualifier of the scope return, thereby preventing any resource in the scoped arena from leaking.

Scopes may nest to form a stack discipline that mirrors classic region-based resource management schemes.
At the meta level, the extent of each scope can be derived from the placement of scoped references, including those in function calls, and we sketch such a translation in \Cref{sec:scope-surface-syntax}.
\section{\arenalang-Calculus: Arena-Based Reachability Tracking} \label{sec:formal}

We present the \arenalang-calculus, an extension of the \weipolylang-calculus \cite{wei24} with shadow arenas, coarse-grained reachability tracking, and two-dimensional store semantics. 
In this section, we primarily focus on the static typing, the call-by-value operational semantics, and the store model.
Dynamic typing is deferred to \Cref{sec:scope-formal}.

\subsection{Syntax} \label{sec:formal-syntax}

\begin{figure}[t]\small
\begin{mdframed}
\begin{minipage}[t]{1.0\textwidth}
\judgement{Syntax}{\BOX{\arenalang}}\small %
  \[\begin{array}{l@{\qquad}l@{\qquad}l@{\qquad}l}
    x,y,z   & \in & \Var        & \text{Variables}               \\
    f,g,h   & \in & \Var       & \text{Function Variables}      \\
    X       & \in & \Var       & \text{Type Variables} \\
    t       & ::= & c \mid x \mid \lambda f(x).t \mid t~t \mid \tref~t \mid \arenacontent{\trefat{t}{t}} \mid\ !~t \mid t \coloneqq t &  \\
            &     & \mid \TLam{X}{x}{t} \mid \TApp{t}{Q}    &  \text{Terms}           \\ [2ex]
    p,q,r,w & \in & \mathcal{P}_{\mathsf{fin}}(\Var \uplus \{ \QFresh \})                           & \text{Reachability Qualifiers} \\
    S,T,U,V & ::= & B \mid f(x: \ty{Q}) \to \ty{Q} \mid \TRef~[Q]   & \\
            &     & \mid \TTop \mid \forall f(\ty[x]{X} <: Q). Q & \text{Types}                   \\
    O,P,Q,R & ::= & \ty[q]{T}                                                                       & \text{Qualified Types}         \\ [2ex]
    \flt    & \in & \mathcal{P}_{\mathsf{fin}}(\Var)                                                & \text{Observations}            \\
    \Gamma  & ::= & \varnothing\mid \Gamma, x : Q  \mid \Gamma, \ty[x]{X} <: Q                                 & \text{Typing Environments}     \\[2ex]
    \end{array}\]

\judgement{Qualifier Notations}
\[ p,q := p \qlub q\qquad x := \{x\} \qquad \QFresh :=\{\QFresh\}\qquad \starred{q} := \{\qfresh\}\qlub q %
\]

\end{minipage} \\[1em]

\caption{The syntax of \arenalang. Extensions from \citet{wei24} and \citet{david} are \arenacontent{\text{shaded}}.}\label{fig:formal-syntax}
\end{mdframed}
\vspace{-2ex}
\end{figure}

\Cref{fig:formal-syntax} defines the formal syntax of \arenalang. 
It follows the conventions of \weipolylang \cite{wei24} for continuity.
Extensions related to shadow arenas are specifically \arenacontent{\text{marked}} to highlight.

Terms include value constants, variables, $\lambda$-abstractions, applications, reference operations, and System \Fsub type abstractions and instantiations. 
Function and type abstractions are explicitly annotated with their self-references to allow their bodies to observe their own reachability.

In addition to standard reference allocation, we introduce \emph{coallocation} of form $\trefat{t_1}{t_2}$ (surface syntax @new Ref(t1) at t2@), which allocates a reference within the arena hosting $t_2$.
Reference operations are uniform across allocation forms, including the scoped allocation form later.

Qualified types $\ty[q]{T}$ are types annotated with reachability qualifiers $q$,
which are finite sets of variables (and later, store locations), and optionally with a freshness marker \QFresh. 
We use $\flt$ to denote an observation filter, a special qualifier specifying the upper bound of resources visible to the typing judgment.
$\Gamma$ denotes the typing context, and later $\Sigma$ denotes the store typing.

The \Fsub-style quantification $\ty[x]{X} <: Q$ introduces a type variable $X$ and a qualifier variable $x$ that can be used separately.
The syntax is an abbreviation for simplicity since types and qualifiers are often quantified together, \ie, $\forall (X <: T).\forall (x <: q).Q \equiv \forall (\ty[x]{X} <: \ty[q]{T}).Q$.

\begin{figure}[t]
\begin{mdframed}
\small
\judgement{Qualifier Substitution}{\BOX{q[p/x]}\ \BOX{q[p/\qfresh]}}
  $$
  \begin{array}{l@{\;}c@{\;}ll@{\quad\qquad\qquad\qquad}l@{\;}c@{\;}ll}
    q[p/x]       &=& q\setminus\{x\}\qlub p & x\in q       & q[p/x]       &=& q & x\notin q  \\
    q[p/\qfresh] &=& q\qlub p               & \qfresh\in q & q[p/\qfresh] &=& q & \qfresh\notin q \\
  \end{array}$$
\judgement{Transitive Closure and Overlap}{\BOX{{\color{gray}\G\vdash}\,\qtrans[n]{q}}\ \BOX{{\color{gray}\G\vdash}\, p \overlap q}}
  $$
\begin{array}{ll@{\quad\qquad}ll}
   \text{Transitive Closure} & \G\vdash\,\qtrans[0]{q} := q  &  \G\vdash\,\qtrans[n+1]{q} := \qtrans[n]{((\bigcup_{x \in q}\, \left\{\, y \mid x : \ty[d,y]{T}\in\G\, \right\}) \cup q)} & \\[1.1ex]
  \text{Qualifier Overlap} & \G\vdash\,p \overlap q := \starred{(\qtrans[\norm{\G}]{p} \qglb\,\, \qtrans[\norm{\G}]{q})}
  & \phantom{some space!}
\end{array}$$

\caption{Computable transitive closure and overlap of qualifiers in $\arenalang$.} \label{fig:formal-qualops}
\end{mdframed}

\vspace{-2ex}
\end{figure} 
\Cref{fig:formal-qualops} defines qualifier substitution, transitive closure, and overlap.
\emph{Transitive closure} computes the maximal qualifier reachable from variables in the original qualifier by recursively unfolding variables to the associated qualifiers in $\Gamma$.
Because qualifiers are assigned minimally for variables, two syntactically disjoint qualifiers may nevertheless refer to shared store locations via aliasing.
Qualifier overlap is therefore defined over transitive closures rather than set intersection.

\subsection{Static Typing and Subtyping} \label{sec:formal-typing-subtyping}

\begin{figure}[t]
\begin{mdframed}
\setlength{\afterruleskip}{\bigskipamount}
\small
\judgement{Term Typing}{\BOX{\strut\G[\flt] \ts t : \ty{Q}}}\\%
\typicallabel{t-var}
\begin{tabular}{b{.22\linewidth}@{}b{.3\linewidth}@{}b{.46\linewidth}}
  \infrule[\ruledef{t1cst}]{
    c \in B
  }{
    \G[\flt] \ts c : \ty[\qbot]{B}
  }
  &
  \infrule[\ruledef{t1var}]{
    x : \ty[q]{T} \in \G\quad x \in \flt
  }{
    \G[\flt] \ts x : \ty[x]{T}
  }
  &
  \infrule[\ruledef{t1sub}]{
    \G[\flt]\ts t : \ty{Q} \quad  \G\ts\ty{Q} <: \ty[q]{T}\quad q\subq\starred{\flt}
  }{
    \G[\flt]\ts t : \ty[q]{T}
  }
\end{tabular} %
\begin{tabular}{m{.48\linewidth}@{}m{.5\linewidth}}
  \infrule[\ruledef{t1ref}]{
    \G[\flt]\ts t : \ty[q]{T} \qquad \QFresh\notin q
  }{
    \G[\flt]\ts \tref~t : \ty[\QFresh]{(\TRef~[\ty[q]{T}])}
  }
  &
  \infrule[\arenacontent{\text{\ruledef{t1refat}}}]{
    \G[\flt]\ts t_1 : \ty[q]{T}\qquad \QFresh\notin q \\
    \G[\flt]\ts t_2 : \ty[p]{\TRef~[U]}
  }{
    \G[\flt]\ts \trefat{t_1}{t_2} : \ty[p]{(\TRef~[\ty[q]{T}])}
  }
  \\[2em]%
  \infrule[\ruledef{t1get}]{
    \G[\flt]\ts t : \ty[q]{\TRef~ [\ty[p]{T}]} \\ \QFresh\notin p 
    \qquad p\subq\flt
  }{
    \G[\flt]\ts !t : \ty[p]{T}
  }
  &
  \infrule[\ruledef{t1put}]{
    \G[\flt]\ts t_1 : \ty[q]{\TRef~ [\ty[p]{T}] } \\
    \G[\flt]\ts t_2 : \ty[p]{T}\quad
    \QFresh\notin p
  }{
    \G[\flt]\ts t_1 \coloneqq t_2 : \ty[\qbot]{\TUnit}
  }
  \\[2em]%
  \infrule[\ruledef{t1abs}]{
    \cx[q,x,f]{(\G\ ,\ f: \ty{F}\ ,\ x: \ty{P})} \ts t : \ty{Q}\\
    \ty{F} = \ty[q]{\left(f(x: \ty{P}) \to \ty{Q}\right)}\qquad q\subq \flt
  }{
    \G[\flt] \ts \lambda f(x).t : \ty{F}
  }
  &
  \infrule[\ruledef{t1tabs}]{
    \cx[q,x,f]{(\G\ ,\ f: \ty{F}\ ,\ \ty[x]{X} <: \ty{P})} \ts t : \ty{Q} \\
    F = \ty[q]{\left(\TAll{X}{x}{P}{}{Q}{}\right)} \qquad q \subq \varphi
  }{
    \G[\flt] \ts \TLam{X}{x}{t} : F
  }
  \\[2em] %
  \infrule[\ruledef{t1app}]{
    \G[\flt]\ts t_1 : \ty[q]{\left(f(x: \ty[p]{T}) \to \ty{Q}\right)} \\
    \G[\flt]\ts t_2 : \ty[p]{T}\qquad \QFresh\notin p \qquad f\notin\FV(U)  \\ 
    Q = \ty[r]{U}\qquad r\subq\starred{\varphi,x,f}
  }{
    \G[\flt]\ts t_1~t_2 : \ty{Q}[p/x, q/f]
  }
  &
  \infrule[\ruledef{t1tapp}]{
    \G[\flt] \ts t : \ty[q]{\left(\TAll{X}{x}{T}{p}{Q}{}\right)} \\
    p \subseteq \varphi \qquad \QFresh \notin p \qquad  f\notin\FV(U) \\
    Q = \ty[r]{U} \qquad r\subq\starred{\flt},x,f
  }{
    \G[\flt] \ts t [ \ty[p]{T} ] : Q[\ty[p]T/\ty[x]{X}, q/f]
  }
  \\[3em] %
  \infrule[\ruledef{t1appfr}]{
    \G[\flt]\ts t_1 : \ty[q]{\left(f(x: \ty[p\,{\overlap}\, q]{T}) \to Q\right)}\\
    \G[\flt]\ts t_2 : \ty[p]{T}\qquad
    \QFresh \in p \Rightarrow x\notin\FV(U) \\  %
    Q = \ty[r]{U}\qquad r\subq\starred{\varphi,x,f}\qquad  f\notin\FV(U) \\
  }{
    \G[\flt]\ts t_1~t_2 : Q[p/x, q/f]
  }
  &
  \infrule[\ruledef{t1tappfr}]{
    \G[\flt] \ts t : \ty[q]{\left(\TAll{X}{x}{T}{p \overlap q}{Q}{}\right)} \\
    p \subseteq \varphi \qquad
    \QFresh \in p \Rightarrow x\notin\FV(U) \\
    Q = \ty[r]{U} \qquad r\subq\starred{\flt},x,f \qquad f\notin\FV(U) \\
  }{
    \G[\flt] \ts t [ \ty[p]{T} ] : Q[\ty[p]{T}/\ty[x]{X}, q/f]
  }
\end{tabular}
\\[1.5em]
\caption{Typing rules of $\arenalang$. Extensions from \citet{wei24} and \citet{david} are \arenacontent{\text{shaded}}.} \label{fig:formal-typing}
\end{mdframed}
\vspace{-2ex}
\end{figure}

\begin{figure}[t]
\setlength{\afterruleskip}{\bigskipamount}
\small
\begin{mdframed}
\judgement{Subtyping}{\BOX{\strut \G\ts\ty{Q} <: \ty{Q}}
                    \ \BOX{\strut \G\ts\ty{T} <: \ty{T}}
                    \ \BOX{\strut\G \ts q <: q}}\\
\typicallabel{sq-sub}
  \infrule[\ruledef{sqsub}]{
    \G\ts\ty{S} <: \ty{T} \qquad \G\ts p <: q
  }{
    \G\ts\ty[p]{S} <: \ty[q]{T}
  }
\vspace{-1em}
\begin{tabular}{b{.2\linewidth}@{}b{.4\linewidth}@{}b{.39\linewidth}}
  \infrule[\ruledef{srefl}]{
  }{
    \G\ts\ty{T} <: \ty{T}
  }
  &
  \infrule[\ruledef{strans}]{
    \G\ts\ty{T} <: \ty{S} \qquad
    \G\ts\ty{S} <: \ty{U}
  }{
    \G\ts\ty{T} <: \ty{U}
  }
  &
  \infrule[\ruledef{qtrans}]{
    \G\ts p <: q \qquad
    \G\ts q <: r
  }{
    \G\ts p <: r
  }
\end{tabular} \\ %
\begin{tabular}{b{.63\linewidth}@{}b{.36\linewidth}}
  \infrule[\ruledef{stop}]{
  }{
    \G\ts \ty{T} <: \TTop
  }
  &
  \infrule[\ruledef{qsub}]{
    p\subq q\subq \starred{\dom(\G)}
  }{
    \G\ts p <: q
  }
  \\[0.5em] %
  \infrule[\ruledef{stvar}]{
    \ty[x]{X} <: \ty[q]{T} \in\G
  }{
    \G\ts \ty{X} <: \ty{T}
  }
  &
  \infrule[\ruledef{qcong}]{
    \G\ts q_1 <: q_2
  }{
    \G\ts p, q_1 <: p, q_2
  }
  \\[0.5em] %
  \infrule[\ruledef{sref}]{
    \G\ts\ty{S} <: \ty{T} \qquad
    \G\ts\ty{T} <: \ty{S} \qquad q\subq\DOM(\Gamma)
  }{
    \G\ts\ty{\TRef~\ty[q]{S}} <: \ty{\TRef~\ty[q]{T}}
  }
  &
  \infrule[\ruledef{qself}]{
    f : \ty[q]{T}\in\G \qquad \QFresh\notin q
  }{
    \G\ts q,f <: f
  }
  \\[0.5em] %
  \infrule[\ruledef{sfun}]{
    \G\ts\ty{P} <: \ty{O} \\
      \G\, ,\, f : \ty[\QFresh]{(f(x : O)\to Q)}\, ,\, x : \ty{P}\ts \ty{Q} <: \ty{R}
  }{
    \G\ts\ty{f(x: \ty{O}) \to \ty{Q}} <: \ty{f(x: \ty{P}) \to \ty{R}}
  }
  &
  \infrule[\ruledef{qvar}]{
    x : \ty[q]{T}\in\G \qquad \QFresh\notin q
  }{
    \G\ts x <: q
  }
  \\[0.5em] %
  \infrule[\ruledef{sall}]{
    \G\ts\ty{P} <: \ty{O} \\
    \G\, ,\, f : \ty[\QFresh]{(\forall f(\ty[x]{X} <: O). Q)}\, ,\, x : \ty[x]{X} <: P \ts \ty{Q} <: \ty{R}
  }{
    \G\ts\ty{\forall f(\ty[x]{X} <: O). Q} <: \ty{\forall f(\ty[x]{X} <: P). R}
  }
  &
  \infrule[\ruledef{qqvar}]{
    \ty[x]{X} <: \ty[q]{T}\in\G \quad \QFresh\notin q
  }{
    \G\ts x <: q
  }
\end{tabular} \\[1.5em]
\caption{Subtyping rules of $\arenalang$.} \label{fig:formal-subtype}
\end{mdframed}

\vspace{-2ex}
\end{figure} 
\Cref{fig:formal-typing} and \Cref{fig:formal-subtype} present the declarative typing and subtyping rules of \arenalang.
All references share a single reference type, independent of their allocation form.
Unlike \citet{david}, who introduce specialized cyclic reference types with constraints, our system retains full deep dependencies of arguments while supporting cycles, which will be discussed in \Cref{sec:case-fixpoint}.
Shadow arenas do not show up in static typing, simplifying the surface syntax.
In the absence of coallocation \arenacontent{\rulename{t1refat}}, \arenalang collapses to the declarative typing system of \citet{wei24},
since coarse-grained reachability tracking degenerates to fine-grained tracking over arenas of size one.

\paragraph{Subtyping.}
Subtyping is defined over qualifiers \rulefmt{q-}, types \rulefmt{s-}, and qualified types \rulename{sqsub}.
Type-level subtyping rules resemble those of System \Fsub and concern qualified
types.
Qualifier subtyping includes context-independent rules \rulename{qsub} and \rulename{qcong} that structurally enlarge qualifiers,
and context-dependent rules \rulename{qvar} and \rulename{qself}. 
Variables may unfold to their recorded qualifiers \rulename{qvar}, and function self-references may additionally fold their captured resources \rulename{qself}.

\paragraph{Term Typing.}
Term typing judgments are regulated by an observation filter $\flt$, which bounds maximal observable resources during typing.
The variable typing rule \rulename{t1var} assigns the variable itself as the minimal qualifier.
Function abstraction \rulename{t1abs} types the body under its function qualifier, argument, and self-reference as the observation filter, ensuring all the resources required for function computation are captured in the function qualifier.

Application is divided into two cases: precise application \rulename{t1app} and growing application \rulename{t1appfr}. 
Precise application requires the argument qualifier to be fully contained in the function's domain qualifier.
Growing application, with a freshness marker \QFresh\  in the function domain qualifier, permits the argument to reach contextually fresh resources beyond the function qualifier.
The function domain qualifier in the growing application is overridden into the upper bound of \emph{controlled sharing}, 
requiring that the overlap between the function qualifier and the argument qualifier ($p\overlap q$) lies within that bound. 
Our application adheres to the same constraints in \citet{wei24}.

The function codomain may depend on function and argument qualifiers by allowing the \emph{deep occurrences} of the bound variables $f$ and $x$ in the return type and qualifier, which supports lightweight reachability polymorphism \cite{wei24}.
\Fsub-style polymorphism is encoded by leveraging the reachability polymorphism. 
The universal instantiation is treated as application to a phantom argument serving as the type witness, so that rules \rulename{t1tabs}, \rulename{t1tapp}, and \rulename{t1tappfr} are structurally similar to function abstraction and applications.

\paragraph{Reference Typing.} 
As references introduced in different forms share a single reference type, reference assignment \rulename{t1put} and dereferencing \rulename{t1get} are generic.
Fresh allocation \rulename{t1ref} introduces a new reference marked fresh, corresponding to standard allocation in \citet{wei24}.
The referent qualifier must be non-fresh to retain sound reachability tracking; otherwise, the same fresh referent could be bound to multiple non-aliasing variables,
leading to unsoundness:
\begin{lstlisting}
  val r  = new Ref(new Ref(42))   // : Ref[Ref[Int]$\trackfresh$]$\trackvar{r}$
  val v1 = !r                     // : Ref[Int]$\trackvar{v1}$
  val v2 = !r                     // : Ref[Int]$\trackvar{v2}$
  free(v1); use(v2)               // $\color{red}{\texttt{Type check but actually use-after-free!}}$
\end{lstlisting}
While the inner qualifier represents the referent reachability, the outer qualifier tracks only the immediate reference reachability, as \emph{shallow reference tracking} \cite{david}.

The freshness marker \QFresh\ in the outer qualifier \rulename{t1ref} conceptually represents a fresh shadow arena with the fresh allocation. 
Subsequent coallocation \rulename{t1refat} places a new reference into an existing arena by assigning it the same outer qualifier as the proxy reference in the @at@ clause, 
as discussed in \Cref{sec:overview-coarse-grained}.
The proxy reference itself may be fresh, \eg, @new Ref(42) at new Ref(7)@ safely identifies an unnamed arena, since the proxy becomes unreachable afterwards.

\subsection{Dynamic Semantics and Two-Dimensional Stores} \label{sec:formal-2dstore}
The \arenalang-calculus generalizes the small-step call-by-value semantics of \weipolylang to a two-dimensional store supporting arena-based memory management. 
\Cref{fig:formal-dynamics} summarizes the operational semantics.

\begin{figure}\small
\begin{mdframed}
\judgement{Term Typing}{\BOX{\cx[\flt]{[\Gamma\mid\Sigma]} \ts t : \ty{Q}}}
\[\ell\in\Loc\qquad \arenacontent{o\in\Offset} \qquad\Sigma ::= \varnothing \mid \Sigma,\arenacontent{\lot{\ell}{o}} : Q \qquad \DOM(\Sigma) := \left\{\lot{\ell}{o}\mid \lot{\ell}{o} \in \Sigma \right\}
\] 
\[ p,q,r \subq \mathcal{P}_{\mathsf{fin}}(\Var \uplus \Loc \uplus \{ \QFresh \})\qquad \flt \subq \mathcal{P}_{\mathsf{fin}}(\Var \uplus \Loc) \qquad \DOML(\Sigma) := \left\{\ell\mid \lot{\ell}{o} \in \Sigma \right\} \]

\infrule[\arenacontent{\text{\ruledef{t1loc}}}]{
    \Sigma(\lo{\ell}{o}) = \ty[{q}]{T}\quad q\subq\DOML(\Sigma)\quad\FV(T)=\varnothing \quad \FTV(T)=\varnothing \quad \ell \in \flt
  }{
    \cx[\flt]{[\G\mid\Sigma]}\ts \lot{\ell}{o} : \ty[\ell]{\TRef~[\ty[q]{T}]}
  }

\judgement{Well-Formed and Well-Typed Stores}{\BOX{\WF{\Sigma}} \BOX{\cx[\flt]{[\G\mid\Sigma]}\ts\sigma}}%
\[
  \cx[\flt]{[\G\mid\Sigma]}\ts \sigma\phantom{m}:=\phantom{m}\flt \subseteq \DOML(\sigma) \subseteq \DOML(\Sigma) \land \forall \ell,\, o,\, \ell \in \flt,\,\lot{\ell}{o}\in\DOM(\Sigma),\, \cx[\flt]{[\G\mid\Sigma]} \ts \sigma\lot{\ell}{o} : \Sigma\lot{\ell}{o}
\]
\vspace{0.5pt}
\[
  \inferrule{\ }{\WF{\varnothing}}\qquad\qquad
  \inferrule{\WF{\Sigma}\quad \FV(T)=\varnothing\quad\FTV(T)=\varnothing \quad q\subseteq\DOML(\Sigma) \quad \arenacontent{o>0\Rightarrow\ell\in\DOML(\Sigma)} }{\WF{\Sigma\, ,\, \lot{\ell}{o} : \ty[q]{T}}}
\]

\vspace{1.5pt}
\judgement{Reduction Contexts, Values, Terms, Stores}{}
\[\begin{array}{l@{\ \ }c@{\ \ }l@{\qquad\qquad\ }l@{\ \ }c@{\ \ }l}
    {C} & ::= & \square \mid C\ t \mid v\ C \mid \tref~C \mid \arenacontent{\trefat{t}{C}} \mid \arenacontent{\trefat{C}{v}} \mid\ !{C} \mid {C} := {t} \mid {v} := {C} \mid C\ [Q] \\
    v &::=& \lambda f(x).t \mid {c} \mid \arenacontent{\lot{\ell}{o}} \mid \tunit \mid \Lambda f(\ty[x]{X}).t  \qquad  
    t ::= \cdots \mid \arenacontent{\lot{\ell}{o}}    \qquad             
    \sigma ::= \varnothing \mid \sigma, \arenacontent{\lot{\ell}{o}}\mapsto v
  \end{array}\]
\judgement{Reduction Rules}{\BOX{t \mid \sigma \to t \mid\sigma}}
\[\begin{array}{r@{\ \ }c@{\ \ }ll@{\qquad\qquad}r}
    \CX[gray]{C}{(\lambda f(x).t)\ v} \mid \sigma     & \to & \CX[gray]{C}{t[v/x, (\lambda f(x).t)/f]} \mid \sigma                  &                           & \rulefmt{$\beta$} \\
    \CX[gray]{C}{\tref~v} \mid \sigma                 & \to & \CX[gray]{C}{\arenacontent{\lot{\ell}{0}}} \mid (\sigma, \arenacontent{\lot{\ell}{0}} \mapsto v)                      & \arenacontent{\lot{\ell}{0}} \not\in \DOM(\sigma) & \rulefmt{ref} \\
    \CX[gray]{C}{\arenacontent{\trefat{v}{\lot{\ell}{o}}}} \mid \sigma                 & \to & \CX[gray]{C}{\arenacontent{\lot{\ell}{o'}}} \mid (\sigma, \arenacontent{\lot{\ell}{o'}} \mapsto v)                      & \lot{\ell}{o'} \not\in \DOM(\sigma)   & \rulefmt{\arenacontent{\text{refat}}} \\
    \CX[gray]{C}{!\lot{\ell}{o}} \mid \sigma                   & \to & \CX[gray]{C}{\sigma(\lot{\ell}{o})} \mid \sigma                                & \lot{\ell}{o}  \in \DOM(\sigma)     & \rulefmt{deref} \\
    \CX[gray]{C}{\lot{\ell}{o} := v} \mid \sigma               & \to & \CX[gray]{C}{\tunit} \mid \sigma[\lot{\ell}{o} \mapsto v]              & \lot{\ell}{o} \in \DOM(\sigma)     & \rulefmt{assign}\\
\CX[gray]{C}{(\Lambda f(\ty[x]{X}).t)\ Q} \mid \sigma & \to &  \CX[gray]{C}{t[Q/\ty[x]{X}, (\Lambda f(\ty[x]{X}).t)/f]} \mid \sigma &                           & \rulefmt{$\beta_T$} \\
  \end{array}\]
\caption{Store typings and call-by-value reduction of \arenalang (\Cref{sec:formal}). Extensions from \citet{wei24} and \citet{david} are \arenacontent{\text{shaded}}. }\label{fig:formal-dynamics}
\end{mdframed}
\vspace{-2ex}
\end{figure}

\subsubsection{Two-Dimensional Stores.}
As illustrated in \Cref{fig:intro-mm-layout}, our two-dimensional store follows the traditional arena-based layouts, mapping from \emph{store indexes}, pairs of \emph{locations} and \emph{offsets} $\lot{\ell}{o}$, to reference cells.
An arena at location $\ell$ conceptually describes the collection of reference cells sharing the same location $\ell$ and differing only in offsets.
The layout is described as \emph{two-dimensional} to emphasize that references are grouped by arenas, while no separation or hierarchy constraints are imposed across arenas.
References may freely reach other arenas regardless of allocation order, in contrast to traditional region-based systems with explicit regions (\Cref{sec:store-topo}).

\subsubsection{Reduction.}
The \arenalang-calculus generalizes the call-by-value reduction of $\weipolylang$ to populate the two-dimensional store.
Fresh allocation and coallocation extend the two-dimensional store in either rows or columns, respectively.
Reducing a coallocation, $\trefat{t_1}{t_2}$, first evaluates the proxy reference $t_2$ to a concrete store index, and the new reference cell is then placed at a fresh offset within that location after the evaluation of $t_1$.
We omit the dynamic typing, as a subset of \Cref{fig:formal-scope-typing}.

Once reduced to a precise store index, a reference is tracked solely by its location $\ell$, equivalently the arena location.
Thus, during reduction, the freshness marker $\QFresh$ in the qualifier is substituted with its location $\ell$, which is reflected in the preservation theorem.

\subsection{Metatheory}

We state the key soundness theorems of \arenalang. 
Proof details are omitted as \arenalang-calculus is a strict subset of the \withlang-calculus (\Cref{sec:formal-scope-theory}).

\begin{theorem}[Progress] \label{theorem:arena-progress}
  If \phantom{n}$\cx[\flt]{[\varnothing\mid\Sigma]}\ts t:Q$ and $\WF{\Sigma}$, then either $t$ is a value, or for any store~$\sigma$ where $\cx[\flt]{[\varnothing\mid\Sigma]}\ts \sigma$, there exists a term $t'$ and a store $\sigma'$ such that $t\mid\sigma \to t' \mid \sigma'$.
\end{theorem}

\begin{theorem}[Preservation] \label{theorem:arena-preservation}
  If \phantom{n}$\cx[\flt]{[\varnothing\mid\Sigma]}\ts t:\ty[q]{T}$, and $\cx[\flt]{[\varnothing\mid\Sigma]}\ts \sigma$, and $\WF{\Sigma}$, and $t\mid\sigma \to t'\mid\sigma'$, then there exists $\Sigma'\supseteq \Sigma, \flt'\supseteq \flt \cup p$, and $p \supseteq \DOML(\Sigma'/\Sigma)$ such that $\cx[\flt']{[\varnothing\mid\Sigma']}\ts\sigma'$ and $\cx[\flt']{[\varnothing\mid\Sigma']}\ts t' : \ty[{ q[p/\qfresh] }]{T}$.  
\end{theorem}

These theorems have been mechanized in Rocq.

\section{\withlang-Calculus: Scoped Resource Management }  \label{sec:scope-formal}

This section presents the \withlang-calculus, which extends the \arenalang-calculus with scoped resource management and deallocation reasoning.
Since our deallocation reasoning requires no type distinctions or surface level annotations, most static typing rules remain unchanged from \arenalang-calculus.
Content specific to dynamic typing is highlighted in \scopedynamicstext{violet}, while removing these dynamic constraints yields the corresponding static system.

\subsection{Language Extensions on \arenalang} \label{sec:formal-scope-static}
\Cref{fig:formal-with-statics} summarizes the language extensions from \arenalang-calculus to \withlang-calculus.
The term syntax is extended with the static scope introduction form, $\twithr{t_1}{t_2}$, and its dynamic elimination form, $\twithc{\ell}{t}$.
The introduction form declares a scoped reference of $t_1$ whose lifetime is lexically bound with $x$ in $t_2$, while the elimination form is internal and not part of the surface syntax.

\paragraph{Static Typing.}
The static typing rule \rulename{t1withr} combines aspects of fresh reference allocation, function abstraction, and application. 
Intuitively, the introduction form $\twithr{t_1}{t_2}$ types the body $t_2$ in a context extended with a freshly allocated reference $x$ for $t_1$, analogous to introducing a reference via let-binding.
The rule follows the idea of scoped acquisition and release discipline, that resources are acquired at scope entry and released at scope exit.
The non-escaping requirement, $x\notin \FV (Q)$, is encoded as deep disjointedness between the scoped reference $x$ and the qualified type $Q$ of the scope return $t_2$, which ensures no downstream dependencies on $x$.

\paragraph{Reduction.}
Reduction of a scope proceeds in two phases.
First, after the resource $t_1$ reduces to a value, a fresh reference (and its underlying arena) is allocated at $\lot{\ell}{0}$ for a fresh $\ell$, mirroring fresh allocation.
The scope introduction form then reduces to its elimination form $\twithc{\ell}{t_2}$, 
where occurrences of the bound variable $x$ in $t_2$ are substituted with the concrete store index $\lot{\ell}{0}$.
Subsequently, $t_2$ is evaluated within the elimination form, 
and once $t_2$ reduces to a value $v$, the elimination form deallocates the entire arena at $\ell$ and unpacks $v$, 
thereby closing the scope.

\begin{figure}\small
\begin{mdframed}
\judgement{Term Syntax, Reduction Contexts}{\BOX{\withlang}}
\[\begin{array}{l@{\ \ }c@{\ \ }l@{\qquad\qquad\ }l@{\ \ }c@{\ \ }l}
  t & ::= & \cdots \mid \twithr{t}{t} \mid \scopedynamics{\twithc{\ell}{t}} \qquad
  {C} ::= \cdots \mid \twithr{C}{t} \mid \twithc{\ell}{C} \\
  \end{array}
\]

\judgement{Reduction Rules}{\BOX{t \mid \sigma \to t \mid\sigma}}
\[\begin{array}{r@{\ \ }c@{\ \ }ll@{\quad}r}
    \CX[gray]{C}{\twithr{v}{t}} \mid \sigma     & \to & \CX[gray]{C}{\twithc{\ell}{t[\lot{\ell}{0}/x]}} \mid (\sigma, \lot{\ell}{0}\mapsto v)   & \lot{\ell}{0} \not\in\DOM(\sigma)  & \rulefmt{$\text{with}$} \\
    \CX[gray]{C}{\twithc{\ell}{v}} \mid \sigma  & \to & \CX[gray]{C}{v} \mid \sigma [\lot{\ell}{*} \mapsto \None] &  & \rulefmt{$\text{close}$}
  \end{array}\]

\judgement{Static Term Typing}{\BOX{\G \ts t:Q}}
  \infrule[\ruledef{t1withr}]{
    \cx[\flt]{\G} \ts t_1 : \ty[p]{T} \quad \cx[\flt,x]{(\G\ ,\ x: \ty[\QFresh]{\TRef~ [\ty[p]{T}]})} \ts t_2 : \ty[]{Q} \quad \QFresh \notin p \quad x \notin \FV(Q) \quad
  }{
    \cx[\flt]{\G} \ts \twithr{t_1}{t_2} : Q
  }

\caption{Extensions of \withlang on \arenalang in syntax, static typing, and reduction. The term $\twithc{\ell}{t}$ (\scopedynamicstext{marked in violet}) is hidden from the surface syntax. }\label{fig:formal-with-statics}
\end{mdframed}
\vspace{-2ex}
\end{figure}

\begin{figure}\small
\begin{mdframed}
\judgement{Selected Syntax Translation}{\BOX{\transform{\sourcecode{t_s}} = \lambda\continuation.\,\targetcode{t_c} } }%
\[\begin{array}{r@{\ \ }c@{\ \ }ll@{\qquad\quad}r}
  \transformsource{new Ref(t)$\,$ scoped} &=& \lambda\continuation.\,\transformsource{t} ~ (\lambda t^\prime.\, \targetcode{\twithr[\cpsfresh{x}]{t^\prime}{\continuationof{ \targetcode{\cpsfresh{x}} }}})  & \cpsfresh{x} \text{ is fresh} & \\
  \transformsource{def f(x) = t} &=& \lambda\continuation.\,\continuationof{ \targetcode{\lambda f(x).}\, \transformsource{t} ~ (\lambda t^\prime.\, \targetcode{t^\prime}) }  & & \\
  \transformsource{t1(t2)} &=& \lambda\continuation.\,\transformsource{t1}~ (\lambda {t_1}'.\, \transformsource{t2} ~ (\lambda {t_2}'.\, \continuationof{\targetcode{{t_1}' ~ {t_2}'}})) & & \\
  \transformsource{x} &=& \lambda \continuation.\,\continuationof{\targetcode{x}}     & & 
\end{array}\]

\caption{The selected syntax translation from the {\color{dark-cyan}surface syntax} to the {\color{magenta}core syntax} of \withlang. }\label{fig:scope-syntax-translation}
\end{mdframed}
\vspace{-2ex}
\end{figure}

\subsection{Surface Syntax Translation}    \label{sec:scope-surface-syntax}
Careful readers may notice that the surface syntax of scope introduction (\Cref{fig:intro-mm-snippet}) does not strictly match the core syntax in the formal calculus.
In particular, scoped allocation (marked with @scoped@) may appear mid-block, whereas the core calculus expects it at block start. 
This discrepancy can be resolved by a translation to align the lifetime of scoped references with blocks.
Namely, a scoped reference declared within a block is translated so that its lifetime extends to the block end.

A subtle case arises when scoped allocation appears as a function argument, \eg, @f(new Ref(t)scoped)@. 
Here, the scoped reference must outlive argument evaluation and potentially align with the lifetime of the function return.
We handle this by a meta-level translation that lifts the allocation outward and binds it earlier.

\Cref{fig:scope-syntax-translation} sketches the translation from surface syntax (in {\color{dark-cyan} dark cyan}) to the core syntax in the formal calculus (in {\color{magenta} magenta}).
The sketch follows the notation convention of \citet{DBLP:conf/lfp/DanvyF90}, where an abstract continuation $\continuation$ is a context-dependent meta-level function passed throughout the translation.
To initiate a translation, we apply $\transform{\sourcecode{t_s}}$ on an identity continuation $\lambda x.\targetcode{x}$.

We showcase translation from the example surface syntax \sourcecode{\lstinline|f(new Ref(v)$\,$ scoped)|} to core calculus:
\[
\setlength{\abovedisplayskip}{2pt}
\setlength{\belowdisplayskip}{2pt}
\begin{array}{r@{\ \ }c@{\ \ }ll@{\qquad\quad}r}
&& \transformsource{f(new Ref(t)$\,$ scoped)} ~ (\lambda x.\,\targetcode{x})    &&      \\
&=& \lambda\continuation.\, \transformsource{f} ~ (\lambda f'.\,\transformsource{new Ref(t)$\,$ scoped}~ (\lambda r'.\,\continuationof{\targetcode{f' ~ r'}})) ~ (\lambda x.\,\targetcode{x})  &\text{application}& \\
&=& (\lambda \continuation.\,\continuationof{\targetcode{f}}) ~ (\lambda f'.\,\transformsource{new Ref(t)$\,$ scoped}~ (\lambda r'.\,\targetcode{f' ~ r'}))  &\text{variable }f& \\
&=& (\lambda\continuation.\, \transformsource{t} ~ (\lambda t'.\, \targetcode{\twithr[\cpsfresh{x}]{t^\prime}{\continuationof{ \targetcode{\cpsfresh{x}} }}})) ~ (\lambda r'.\,\targetcode{f ~ r'})   &\text{scoped reference}& \\
&=& (\lambda\continuation.\, \continuationof{\targetcode{t}}) ~ (\lambda t'.\, \targetcode{\twithr[\cpsfresh{x}]{t^\prime}{f ~ \cpsfresh{x}}})  &\text{variable }t& \\
&=& \targetcode{\twithr[\cpsfresh{x}]{t}{\targetcode{f ~ \cpsfresh{x}} } } & & 
\end{array}\]

After translation, the function $\targetcode{f}$ is applied to the scoped binding~$\targetcode{\cpsfresh{x}}$ within the scope.
Other translation rules can be defined straightforwardly and omitted.

\subsection{Flow-Insensitive Deallocation Reasoning} \label{sec:scope-reasoning}
The design principle of \withlang-calculus is to introduce lexical scopes following stack discipline without requiring additional surface annotations or type distinctions.
Rather than relying on explicit use-and-kill effects, the calculus reasons about safe deallocation through complementary forms of observation filters and dynamic tracking of locations to be reclaimed.
The \emph{flow-insensitive} deallocation reasoning is established by dynamic typing and preserved under reduction without being exposed at the static level.
We start from discussing why static non-escaping is sufficient for deallocation safety, and then illustrate how the property is preserved dynamically.

\subsubsection{Non-Escaping in Static Typing.}
The typing rule \rulename{t1withr} enforces that the scoped variable $x$ does not occur in the type or qualifier of the scope return $t_2$. 
This ensures that $x$ is disjoint from downstream computation, since any transitive reachability to $x$ via store lookup would induce a deep occurrence of $x$ in the type:
\begin{lstlisting}
  val x = new Ref(..) scoped      // : Ref[T]$\trackvar{x}$ $\dashv$ [ x : Ref[T]$\trackfresh$ ], scope body starts
  val r1 = x                      // : Ref[T]$\trackvar{r1}$ <: Ref[T]$\trackvar{x}$, $\color{red}{\texttt{cannot return r1}}$
  val r2 = new Ref(x)             // : Ref[Ref[T]$\trackvar{x}$]$\trackvar{r2}$, $\color{red}{\texttt{cannot return r2}}$
  val r3 = new Ref(x)             // : $\TTop\trackvar{r3}$, OK to return
  val r4 = new Ref(!x)            // : Ref[T]$\trackvar{r4}$, OK to return
\end{lstlisting}
Because the scope return must be typed without observing $x$, internal aliases or references such as @r1@ and @r2@ are rejected. 
By contrast, @r3@ is permitted after being upcasted to $\TTop$, which prevents dereferencing and thus guarantees that $x$ is unreachable. 
Our system guarantees the reclamation of $x$ itself, but not of its referent, so returning a disjoint reference such as @r4@ is allowed.

\subsubsection{Non-Trivial Gaps in Dynamics.}
While the non-escaping property prevents scoped references from escaping their scope,
it does not prevent reintroduction of reachability to reclaimed resources during later evaluation.
Statically, reintroduction is impossible because the variable binding $x$ is only appended to $\Gamma$ when typing the body $t_2$ and not observable otherwise. 
Dynamically, however, store locations of scoped references can be accessed directly \rulename{t1loc}, bypassing the $\G$ context. 

Recall the preservation theorem (\Cref{theorem:arena-preservation}) inherited from prior reachability type systems \cite{david,wei24}:
the observation filter $\flt$ monotonically grows to $\flt \cup p$ where $p$ contains fresh store locations from the reduction.
As a result, a reclaimed location could be reintroduced through subsumption \rulename{t1sub} or growing application \rulename{t1appfr}, 
unless it is explicitly removed from observation $\flt$. 
This evidently illustrates the central challenge of deallocation safety.

\subsubsection{Observation Filters.}
To prevent reintroduction of reclaimed locations, deallocated locations must be excluded from the observation filter $\flt$.
This entails the invariant that $\flt$ remains disjoint from all scoped locations outside their lexical scopes.
Accordingly, for a scoped elimination form $\twithc{\ell}{t_2}$: 
(1) $\ell$ is added to $\flt$ when the elimination form is introduced; 
(2) $\ell$ remains observable during the evaluation of $t_2$; and
(3) $\ell$ is removed from $\flt$ when the evaluation finishes.

However, shrinking the observation filter $\flt$ cannot be arbitrary, as \emph{strengthening} the typing judgment with fewer observable resources is not generally admissible; that is,
$\cx[\flt]{[\G\mid\Sigma]}\ts t : T \centernot\implies \cx[\flt\setminus\ell]{[\G\mid\Sigma]}\ts t : T$.
To address this, observation filters are shrunk only when a scope elimination form reduces to its value (\Cref{sec:scope-strengthening}),
and the dynamic typing ensures that such shrinking does not affect typing outside that scope (\Cref{sec:scope-local-locs-typing}). 
This is achieved via a combination of \emph{local locations} and \emph{well-formed stepping} (\Cref{fig:formal-with-dynamics}), as well as dynamic typing rules (\Cref{fig:formal-scope-typing}).

\subsubsection{Strengthening with Value Witnesses.} \label{sec:scope-strengthening}
In reachability types, the qualifier of a \emph{value} characterizes the minimal resources required to type it. 
For example, the body of a function abstraction \rulename{t1abs} is typed with observation limited to the function qualifier $q$ instead of $\flt$.
Formally, qualifiers assigned to well-typed values provide lower bounds for observation in typing.
\begin{lemma}[Value Observability] \label{theorem:scope-value-observation}
  If $\cx[\flt]{[\G\mid\Sigma]}\ts v:\ty[q]{T}$, then $\cx[q]{[\G\mid\Sigma]}\ts v:\ty[q]{T}$.
\end{lemma}

With a value witness, we can shrink the observation filter down to any $\flt'$ that contains $q$.
\begin{lemma}[Value Retyping] \label{theorem:scope-value-retype}
  If $\cx[\flt]{[\G\mid\Sigma]}\ts v:\ty[q]{T}$, and $\ell\notin q$, then $\cx[\flt\setminus\ell]{[\G\mid\Sigma]}\ts v:\ty[q]{T}$.
\end{lemma}
This lemma justifies removing a scoped location from $\flt$ at scope exit.
Once $t_2$ reduces to a value $v$, the premise $\ell \notin q$ required by the typing rule \rulename{t2withc} 
allows $\ell$ to be safely eliminated from the observation filter without affecting the typing of $v$.

\begin{figure}[t]
\begin{mdframed}
\setlength{\afterruleskip}{\bigskipamount}
\small
\judgement{Dynamic Term Typing}{\BOX{\strut \cx[\flt]{[\G\mid\Sigma]} \ts t : \ty{Q}}}\\[-1ex]
\typicallabel{t-var}
\begin{tabular}{b{.49\linewidth}@{}b{.5\linewidth}}
  \infrule[\ruledef{t2cst}]{
    c \in B
  }{
    \cx[\flt]{[\G\mid\Sigma]} \ts c : \ty[\qbot]{B}
  }
  &
  \infrule[\ruledef{t2var}]{
    x : \ty[q]{T} \in \G\quad\quad x \in \flt
  }{
    \cx[\flt]{[\G\mid\Sigma]} \ts x : \ty[x]{T}
  }
  \\ [0.5em]
  \infrule[\ruledef{t2ref}]{
    \cx[\flt]{[\G\mid\Sigma]}\ts t : \ty[q]{T} \qquad \QFresh\notin q
  }{
    \cx[\flt]{[\G\mid\Sigma]}\ts \tref~t : \ty[\QFresh]{(\TRef~[\ty[q]{T}])}
  }
  &
  \infrule[\ruledef{t2refat}]{
    \cx[\flt\scopedynamics{\ominus t_2}]{[\G\mid\Sigma]}\ts t_1 : \ty[q]{T}\qquad \QFresh\notin q \\
    \cx[\flt]{[\G\mid\Sigma]}\ts t_2 : \ty[p]{\TRef~[U]} \quad \scopedynamics{\scriptstyle\killsep{\LC{t_1}}{p}}
  }{
    \cx[\flt]{[\G\mid\Sigma]}\ts \trefat{t_1}{t_2} : \ty[p]{(\TRef~[\ty[q]{T}])}
  }
  \\[0.5em] %
  \infrule[\ruledef{t2get}]{
    \cx[\flt]{[\G\mid\Sigma]}\ts t : \ty[q]{\TRef~ [\ty[p]{T}]} \\ \QFresh\notin p \qquad
    p\subq\flt \qquad \scopedynamics{\scriptstyle\killsep{\LC{t}}{p}}
  }{
    \cx[\flt]{[\G\mid\Sigma]}\ts !t : \ty[p]{T}
  }
  &
  \infrule[\ruledef{t2put}]{
    \cx[\flt]{[\G\mid\Sigma]}\ts t_1 : \ty[q]{\TRef~ [\ty[p]{T}] } \qquad  \QFresh\notin p \\
    \cx[\flt\scopedynamics{\ominus t_1}]{[\G\mid\Sigma]}\ts t_2 : \ty[p]{T}\quad \scopedynamics{\scriptstyle\killsep{\LC{t_2}}{q}} 
  }{
    \cx[\flt]{[\G\mid\Sigma]}\ts t_1 \coloneqq t_2 : \ty[\qbot]{\TUnit}
  }
  \\[0.5em] %
  \infrule[\ruledef{t2abs}]{
    \cx[q,x,f]{[\G\ ,\ f: \ty{F}\ ,\ x: \ty{P}\mid\Sigma]} \ts t : \ty{Q}\\
    \ty{F} = \ty[q]{\left(f(x: \ty{P}) \to \ty{Q}\right)}\quad q\subq \flt \quad \scopedynamics{\scriptstyle\killsep{\LC{t}}{q}}
  }{
    \cx[\flt]{[\G\mid\Sigma]} \ts \lambda f(x).t : \ty{F}
  }
  &
  \infrule[\ruledef{t2tabs}]{
    \cx[q,x,f]{[\G\ ,\ f: \ty{F}\ ,\ \ty[x]{X} <: \ty{P}\mid\Sigma]} \ts t : \ty{Q} \\
    F = \ty[q]{\left(\TAll{X}{x}{P}{}{Q}{}\right)} \quad q \subq \varphi \quad \scopedynamics{\scriptstyle\killsep{\LC{t}}{q}}
  }{
    \cx[\flt]{[\G\mid\Sigma]} \ts \TLam{X}{x}{t} : F
  }
  \\ [0.5em]
  \infrule[\ruledef{t2app}]{
    \cx[\flt]{[\G\mid\Sigma]}\ts t_1 : \ty[q]{\left(f(x: \ty[p]{T}) \to \ty{Q}\right)}  \quad Q = \ty[r]{U}  \\
    \cx[\flt\scopedynamics{\ominus t_1}]{[\G\mid\Sigma]}\ts t_2 : \ty[p]{T}  \qquad \QFresh\notin p \quad
    f\notin\FV(U)\\ r\subq\starred{\varphi,x,f} \quad
    \scopedynamics{\scriptstyle\killsep{\LC{t_2}}{q,r}} \quad \scopedynamics{\scriptstyle\killsep{\LC{t_1}}{r}} 
  }{
    \cx[\flt]{[\G\mid\Sigma]}\ts t_1~t_2 : \ty{Q}[p/x, q/f]
  }
  &
  \infrule[\ruledef{t2tapp}]{
    \cx[\flt]{[\G\mid\Sigma]} \ts t : \ty[q]{\left(\TAll{X}{x}{T}{p}{Q}{}\right)} \quad Q = \ty[r]{U} \\
    p \subseteq \varphi  \qquad  \QFresh \notin p \qquad f\notin\FV(U) \\
    r\subq\starred{\flt},x,f \qquad \scopedynamics{\scriptstyle\killsep{\LC{t}}{p,r}}
  }{
    \cx[\flt]{[\G\mid\Sigma]} \ts t [ \ty[p]{T} ] : Q[\ty[p]T/\ty[x]{X}, q/f]
  }
  \\ [-1em]
  \infrule[\ruledef{t2appfr}]{
    \cx[\flt]{[\G\mid\Sigma]}\ts t_1 : \ty[q]{\left(f(x: \ty[p\,{\overlap}\, q]{T}) \to Q\right)} \quad Q = \ty[r]{U}  \\
    \cx[\flt\scopedynamics{\ominus t_1}]{[\G\mid\Sigma]}\ts t_2 : \ty[p]{T}  \quad \QFresh \in p \Rightarrow x\notin\FV(U)  \\
    r\subq\starred{\varphi,x,f} \qquad f\notin\FV(U) \\
    \scopedynamics{\scriptstyle\killsep{\LC{t_2}}{q,r}} \qquad \scopedynamics{\scriptstyle\killsep{\LC{t_1}}{(p\,{\overlap}\, q),r}} 
  }{
    \cx[\flt]{[\G\mid\Sigma]}\ts t_1~t_2 : Q[p/x, q/f]
  }
  &
  \infrule[\ruledef{t2tappfr}]{
    \cx[\flt]{[\G\mid\Sigma]} \ts t : \ty[q]{\left(\TAll{X}{x}{T}{p\,{\overlap}\, q}{Q}{}\right)} \quad Q = \ty[r]{U}\\
    p \subseteq \varphi \qquad \QFresh \in p \Rightarrow x\notin\FV(U) \\ r\subq\starred{\flt},x,f \quad f\notin\FV(U) \quad
    \scopedynamics{\scriptstyle\killsep{\LC{t}}{(p\,{\overlap}\, q),p,r}}
  }{
    \cx[\flt]{[\G\mid\Sigma]} \ts t [ \ty[p]{T} ] : Q[\ty[p]{T}/\ty[x]{X}, q/f]
  }
  \\ [-1em]
  \infrule[\ruledef{t2sub}]{
    \cx[\flt]{[\G\mid\Sigma]}\ts t : \ty{Q} \qquad  \G\ts\ty{Q} <: \ty[q]{T}\\
    q\subq\starred{\flt} \qquad \scopedynamics{\scriptstyle\killsep{\LC{t}}{q}}
  }{
    \cx[\flt]{[\G\mid\Sigma]}\ts t : \ty[q]{T}
  }
  &
  \infrule[\ruledef{t2withr}]{
    \cx[\flt]{[\G\mid\Sigma]} \ts t_1 : \ty[p]{T} \quad \QFresh \notin p \quad x \notin \FV(Q) \\
    \cx[\flt\scopedynamics{\ominus t_1},x]{[\G\ ,\ x: \ty[\QFresh]{\TRef~ [\ty[p]{T}]} \mid \Sigma]} \ts t_2 : \ty[]{Q}
  }{
    \cx[\flt]{[\G\mid\Sigma]} \ts \twithr{t_1}{t_2} : Q
  }
  \\ [0.5em]
  \infrule[\scopedynamicstext{\ruledef{t2loc}}]{
    \Sigma(\lo{\ell}{o}) = \ty[{q}]{T}\qquad  \ell \in \flt \qquad \QFresh\notin q
  }{
    \cx[\flt]{[\G\mid\Sigma]}\ts \lot{\ell}{o} : \ty[\ell]{\TRef~[\ty[q]{T}]}
  }
  &
  \infrule[\scopedynamicstext{\ruledef{t2withc}}]{
    \cx[\flt]{[\G\mid\Sigma]} \ts t : \ty[q]{T}  \qquad  \ell \notin q
  }{
    \cx[\flt]{[\G\mid\Sigma]} \ts \twithc{\ell}{t} : \ty[q]{T}
  }
\end{tabular} \\[1.5em]
\caption{Term typing for program dynamics of \withlang with store typing. The content \scopedynamicstext{marked in violet} are hidden and implicitly inferred for a static program with an empty store. } \label{fig:formal-scope-typing}
\end{mdframed}
\vspace{-2ex}
\end{figure}
 
\subsubsection{Local Locations in Dynamic Typing.} \label{sec:scope-local-locs-typing}
The final challenge is to ensure that scoped allocation and deallocation within one subterm do not affect unrelated subterms, up to typing. 
Consider an application concerning scopes, $(\twithr{v}{t_2}) ~ t$, where $v$ is already a value.
Reducing the left subterm introduces a local location $\ell$ in its elimination form, $\twithc{\ell}{t_2}$.
If both subterms were typed under the same observation filter, $\ell$ would incorrectly propagate to the typing of $t$,
requiring general strengthening when $\ell$ is reclaimed before the reduction of $t$, which is impossible.

To address this, dynamic typing (\Cref{fig:formal-scope-typing}) distinguishes \emph{local locations} (\Cref{fig:formal-with-dynamics})---locations that will be reclaimed in the future.
For example in rule \rulename{t2app}, typing of $t_2$ explicitly excludes the local locations of $t_1$ in its observation filter,
which is sound because those locations are guaranteed to be reclaimed before $t_2$ reduces.
Conversely, local locations of $t_2$ do not affect typing of $t_1$, since $t_1$ is already a value, according to the reduction order described by \emph{well-stepped terms} (\Cref{fig:formal-with-dynamics}).

Together, these mechanisms establish deallocation safety: scoped references and their store locations are safely removed from observation filters at scope exit and are never reintroduced during subsequent reduction, ensuring they remain unreachable downstream.

\subsection{Formal Definitions and Invariants}

After \Cref{sec:scope-reasoning} presents the intuition behind deallocation reasoning, 
we now formalize the definitions and invariants.

\begin{figure}[t]
\begin{mdframed}
\small
\judgement{Static Term Typing}{\BOX{\strut \cx[\flt]{\G} \ts t : \ty{Q}}}\\[-1ex]
\typicallabel{ts-var}
\begin{tabular}{b{.48\linewidth}@{}b{.5\linewidth}}
  \infrule[\ruledef{tscst}]{
    c \in B
  }{
    \cx[\flt]{\G} \ts c : \ty[\qbot]{B}
  }
  &
  \infrule[\ruledef{tsvar}]{
    x : \ty[q]{T} \in \G\quad\quad x \in \flt
  }{
    \cx[\flt]{\G} \ts x : \ty[x]{T}
  }
  \\ [0.5em]
  \infrule[\ruledef{tsref}]{
    \cx[\flt]{\G}\ts t : \ty[q]{T} \qquad \QFresh\notin q
  }{
    \cx[\flt]{\G}\ts \tref~t : \ty[\QFresh]{(\TRef~[\ty[q]{T}])}
  }
  &
  \infrule[\ruledef{tsrefat}]{
    \cx[\flt]{\G}\ts t_1 : \ty[q]{T}\qquad \QFresh\notin p,q \\
    \cx[\flt]{\G}\ts t_2 : \ty[p]{\TRef~[U]}
  }{
    \cx[\flt]{\G}\ts \trefat{t_1}{t_2} : \ty[p]{(\TRef~[\ty[q]{T}])}
  }
  \\[0.5em]
  \infrule[\ruledef{tsget}]{
    \cx[\flt]{\G}\ts t : \ty[q]{\TRef~ [\ty[p]{T}]} \\ \QFresh\notin p \qquad
    p\subq\flt
  }{
    \cx[\flt]{\G}\ts !t : \ty[p]{T}
  }
  &
  \infrule[\ruledef{tsput}]{
    \cx[\flt]{\G}\ts t_1 : \ty[q]{\TRef~ [\ty[p]{T}] } \\
    \cx[\flt]{\G}\ts t_2 : \ty[p]{T} \qquad  \QFresh\notin p 
  }{
    \cx[\flt]{\G}\ts t_1 \coloneqq t_2 : \ty[\qbot]{\TUnit}
  }
  \\[0.5em]
  \infrule[\ruledef{tsabs}]{
    \cx[q,x,f]{(\G\ ,\ f: \ty{F}\ ,\ x: \ty{P})} \ts t : \ty{Q}\\
    \ty{F} = \ty[q]{\left(f(x: \ty{P}) \to \ty{Q}\right)}\quad q\subq \flt 
  }{
    \cx[\flt]{\G} \ts \lambda f(x).t : \ty{F}
  }
  &
  \infrule[\ruledef{tstabs}]{
    \cx[q,x,f]{(\G\ ,\ f: \ty{F}\ ,\ \ty[x]{X} <: \ty{P})} \ts t : \ty{Q} \\
    F = \ty[q]{\left(\TAll{X}{x}{P}{}{Q}{}\right)} \quad q \subq \varphi 
  }{
    \cx[\flt]{\G} \ts \TLam{X}{x}{t} : F
  }
  \\[0.5em]
  \infrule[\ruledef{tsapp}]{
    \cx[\flt]{\G}\ts t_1 : \ty[q]{\left(f(x: \ty[p]{T}) \to \ty{Q}\right)}  \\ Q = \ty[r]{U}  \quad
    \cx[\flt]{\G}\ts t_2 : \ty[p]{T}  \quad \QFresh\notin p \\
    f\notin\FV(U) \quad r\subq\starred{\varphi,x,f} 
  }{
    \cx[\flt]{\G}\ts t_1~t_2 : \ty{Q}[p/x, q/f]
  }
  &
  \infrule[\ruledef{tstapp}]{
    \cx[\flt]{\G} \ts t : \ty[q]{\left(\TAll{X}{x}{T}{p}{Q}{}\right)} \\ Q = \ty[r]{U} \quad
    p \subseteq \varphi  \qquad  \QFresh \notin p \\ f\notin\FV(U) \quad
    r\subq\starred{\flt},x,f
  }{
    \cx[\flt]{\G} \ts t [ \ty[p]{T} ] : Q[\ty[p]T/\ty[x]{X}, q/f]
  }
  \\[0.5em]
  \infrule[\ruledef{tsappfr}]{
    \cx[\flt]{\G}\ts t_1 : \ty[q]{\left(f(x: \ty[p\,{\overlap}\, q]{T}) \to Q\right)}  \\
    \cx[\flt]{\G}\ts t_2 : \ty[p]{T}  \quad \QFresh \in p \Rightarrow x\notin\FV(U)  \\
    Q = \ty[r]{U} \quad r\subq\starred{\varphi,x,f} \qquad f\notin\FV(U)
  }{
    \cx[\flt]{\G}\ts t_1~t_2 : Q[p/x, q/f]
  }
  &
  \infrule[\ruledef{tstappfr}]{
    \cx[\flt]{\G} \ts t : \ty[q]{\left(\TAll{X}{x}{T}{p\,{\overlap}\, q}{Q}{}\right)} \\
    p \subseteq \varphi \qquad \QFresh \in p \Rightarrow x\notin\FV(U) \\ 
    Q = \ty[r]{U} \quad r\subq\starred{\flt},x,f \quad f\notin\FV(U) 
  }{
    \cx[\flt]{\G} \ts t [ \ty[p]{T} ] : Q[\ty[p]{T}/\ty[x]{X}, q/f]
  }
  \\ [0.5em]
  \infrule[\ruledef{tssub}]{
    \cx[\flt]{\G}\ts t : \ty{Q}  \\ 
    \G\ts\ty{Q} <: \ty[q]{T} \qquad q\subq\starred{\flt}
  }{
    \cx[\flt]{\G}\ts t : \ty[q]{T}
  }
  &
  \infrule[\ruledef{tswithr}]{
    \cx[\flt]{\G} \ts t_1 : \ty[p]{T} \quad \QFresh \notin p \quad x \notin \FV(Q) \\
    \cx[\flt,x]{(\G\ ,\ x: \ty[\QFresh]{\TRef~ [\ty[p]{T}]})} \ts t_2 : \ty[]{Q}
  }{
    \cx[\flt]{\G} \ts \twithr{t_1}{t_2} : Q
  }\\[2ex]
\end{tabular}
\caption{Static term typing for programs in \withlang with an empty store. } \label{fig:formal-scope-typing-static}
\end{mdframed}
\vspace{-3ex}
\end{figure}
 
\Cref{fig:formal-scope-typing-static} presents the static typing. 
\Cref{fig:formal-with-dynamics} defines the computation of \emph{local locations} and the notion of \emph{well-stepped terms} (\Cref{fig:formal-with-dynamics}),
which constrain how local locations may appear during reduction.

\paragraph{Dynamic Typing.}
\Cref{fig:formal-scope-typing} defines the dynamic typing rules of \withlang-calculus, formalizing the isolation principle described above: local locations of one subterm must not influence the observations or qualifiers of unrelated subterms.
Concretely, each typing rule subtracts the local locations of sibling subterms from the observation filter.

For example, rule \rulename{t2app} types $t_2$ under $\flt$ minus the local locations of $t_1$, so that allocations and deallocations within $t_1$ do not affect the typing of $t_2$.
Once $t_1$ is fully reduced to a value, its local location set is empty by the constraints on \emph{well-stepped terms}, and thus $t_2$ is typed with the original $\flt$.

Since local locations are always excluded from the observation filter, deallocated locations are never reintroduced.

\begin{figure}\small
\begin{mdframed}

\judgement{\scopedynamicstext{Local Locations} and Qualifier Operations}{\BOX{\LC{t} \in \mathcal{P}_{\mathsf{fin}}(\Loc)}}
\[\begin{array}{l}
  \LC{\twithc{\ell}{t}} := \ell, \LC{t}  \qquad \LC{c},\, \LC{x},\, \LC{\lot{\ell}{o}} := \varnothing  \\
  \LC{\lambda f(x).t},\, \LC{\tref~t},\, \LC{!~t},\, \LC{\TLam{X}{x}{t}},\, \LC{\TApp{t}{Q}{} } := \LC{t} \\
  \LC{t_1~t_2},\, \LC{\trefat{t_1}{t_2}},\, \LC{t_1 \coloneqq t_2},\, \LC{\twithr[]{t_1}{t_2}} := \LC{t_1}, \LC{t_2} \\
  q \ominus t := q \setminus \LC{t}
\end{array}\]

\judgement{\scopedynamicstext{Well-Stepped Terms}}{\BOX{\WT{t}}}
\[
  \inferrule{\WT{t} }{\WT{\twithc{\ell}{t}}}   \qquad
  \inferrule{\WT{t_1}\quad \WT{t_2} \quad \LC{t_2} = \varnothing }{\WT{\twithr{t_1}{t_2}}} \qquad
  \inferrule{\WT{t_1}\quad \WT{t_2} \quad \text{value}~t_1 \lor \LC{t_2}=\varnothing}{\WT{t_1~t_2}}
\] \\
\[
  \inferrule{\WT{t}}{\WT{\tref~t}}\qquad
  \inferrule{\WT{t_1}\quad\WT{t_2}\quad \LC{t_1}=\varnothing \lor \text{value}~t_2}{\WT{\trefat{t_1}{t_2}}}\qquad 
  \inferrule{\WT{t_1}\quad \WT{t_2} \quad \text{value}~t_1 \lor \LC{t_2}=\varnothing}{\WT{t_1 \coloneqq t_2}}
\] \\
\[
  \inferrule{\LC{t}=\varnothing}{\WT{\lambda f(x).t}} \qquad
  \inferrule{\WT{t}}{\WT{!~t}} \qquad
  \inferrule{\ }{\WT{\tunit}} \qquad
  \inferrule{\ }{\WT{\lot{\ell}{o}}}\qquad
  \inferrule{\ }{\WT{x}} \qquad
  \inferrule{\WT{t}}{\WT{\TLam{X}{x}{t}}} \qquad
  \inferrule{\WT{t}}{\WT{t\ [\ty{Q}]}}
\]

\caption{The local location computation and well-stepped terms in \withlang-calculus. }\label{fig:formal-with-dynamics}
\end{mdframed}
\vspace{-1ex}
\end{figure}

\begin{figure}\small
\begin{mdframed}
\judgement{Environment Update}{\BOX{\kappa} \BOX{\Sigma,\flt,\kappa \envto{t} \Sigma',\flt',\kappa'}}%
\[
  \kappa \in \mathcal{P}_{\mathsf{fin}}(\Loc) \qquad  %
  \qquad   \text{Already-Killed Locations}
\]
\[ 
  \inferrule[\ruledef{eubase}]{\DOML(\Sigma)=\DOML(\Sigma')}{\Sigma,\flt,\kappa\envto{t} \Sigma',\flt,\kappa} \qquad
  \inferrule[\ruledef{eufresh}]{\killsep{\kappa}{q}}{\Sigma,\flt,\kappa\envto{t} (\Sigma,\,\lot{\ell}{0} : \ty[q]{T}), (\flt,\,\ell), \kappa} \qquad 
  \inferrule[\ruledef{eukill}]{\ell\in\DOML(\Sigma)\quad \ell\in\LC{t}}{\Sigma,\flt,\kappa\envto{t}\Sigma, (\flt\setminus\ell), (\kappa,\,\ell)}
\]

\vspace{2ex}
\judgement{Well-Formed and Well-Typed Stores}{\BOX{\WF{\Sigma}} \BOX{\cx[\flt]{[\G\mid\Sigma]}\ts\sigma\mid\kappa}}
\[
\begin{array}{l@{\ \ }l}
  \cx[\flt]{[\G\mid\Sigma]}\ts \sigma\mid\kappa :=&  \flt \subseteq \DOML(\sigma) \subseteq \DOML(\Sigma) \land \killsep{\flt}{\kappa} \,\land \\
  &\forall \ell \in \kappa,\, \forall o,\,\lot{\ell}{o}\in\DOM(\sigma),\,\sigma\lot{\ell}{o}=\None \, \land \\
  &\forall \ell \in \flt,\,\forall o,\, \lot{\ell}{o}\in\DOM(\sigma),\,\exists v,\, \sigma\lot{\ell}{o}=v \, \land \WT{v}\, \land \\ 
  & \qquad\quad\, \forall T,q,\, \Sigma\lot{\ell}{o}=\ty[q]{T},\, q\subseteq\flt \to \cx[\flt]{[\G\mid\Sigma]} \ts \sigma\lot{\ell}{o} : \Sigma\lot{\ell}{o}
\end{array}
\]
\vspace{1pt}
\[
  \inferrule{\ }{\WF{\varnothing}}\qquad\qquad
  \inferrule{\WF{\Sigma}\quad \FV(T)=\varnothing\quad\FTV(T)=\varnothing \quad q\in\DOML(\Sigma) \quad o>0\Rightarrow\ell\in\DOML(\Sigma)}{\WF{\Sigma\, ,\, \lot{\ell}{o} : \ty[q]{T}}}
\]
\caption{The environment invariants of the \withlang-calculus. }\label{fig:formal-with-wfstore}
\end{mdframed}
\vspace{-2ex}
\end{figure} 
\paragraph{Environment Update.}
\Cref{fig:formal-with-wfstore} formalizes environment evolution under reduction, refining the well-formedness of \arenalang-calculus.
We introduce a meta-level variable $\kappa$ recording reclaimed locations, which is derivable from the store state.
Unlike prior reachability type systems, which can arbitrarily choose larger context filters for preservation, we strictly specify how observation filters evolve with respect to the store and $\kappa$.
We distinguish three cases:
\begin{enumerate}[leftmargin=1.8em,itemsep=4pt]
  \item \rulename{eufresh}: Allocation of a fresh arena, either permanently or locally, grows the observation $\flt$.
  \item \rulename{eukill}: Deallocation of a local arena shrinks $\flt$ and extends $\kappa$ with the same location. Only local locations captured by the pre-reduction term can be freed.
  \item \rulename{eubase}: Other reductions that do not change the observation $\flt$. Coallocation also falls into this category.
\end{enumerate}

This is necessary because of the invariant for safe deallocation that the observation filter $\flt$ must remain disjoint from killed locations $\kappa$. 
Formalizing the environment update refines the proof structure for deriving more complicated results; however, it introduces notable extra engineering complexity compared with prior work, and we refer interested readers to our artifact for details.

\paragraph{Store Invariants.}
In \Cref{fig:formal-with-wfstore}, the store invariants describe well-formed store typing, reachability tracking, and deallocation.
The well-formed store $\WF{\Sigma}$ captures the \emph{closedness} of each cell in the store typing: the type of each store object must be derivable from existing resources at the point it is appended to $\Sigma$. 
Because the store is two-dimensional, it does not naturally follow the telescoping structure as in prior reachability type systems \cite{wei24, david}, resulting in a weaker and more general store invariant.

The well-typed store judgment, $\cx[\flt]{[\G\mid\Sigma]}\ts\sigma\mid\kappa$, is the most critical invariant for type and store safety. 
For a store $\sigma$, the well-typed store requires:
(1) the observation filter $\flt$ is closed and disjoint from $\kappa$; 
(2) locations in $\kappa$ are reclaimed; and
(3) objects having locations in $\flt$ are typeable with the types recorded in the store typing $\Sigma$.
Crucially, the disjointedness between $\kappa$ and $\flt$ guarantees that well-typed terms cannot observe deallocated resources.
We only require store locations in $\flt$ to agree with the store typing $\Sigma$, as the remainder are either reclaimed or irrelevant to typing.

\subsection{Metatheory} \label{sec:formal-scope-theory}

We establish syntactic type soundness for the \withlang-calculus via progress and preservation theorems.
Our proof structure follows prior work \cite{wei24,david}, with lemma statements and definitions refined to account for deallocation reasoning.

\subsubsection{Observability Properties.}
The observation is the key mechanism to reason about the resources necessary to type a term. 
Reasoning about the separation between observation and killed locations can implicitly guarantee the absence of killed locations in the term typing without exposing the $\kappa$.
The following lemmas are proved by induction over the term typing.

\begin{lemma}[Observability]
  Term typing cannot assign qualifiers beyond observation, \ie, if $\cx[\flt]{[\G\mid\Sigma]}\ts t:\ty[q]{T}$, then $q\subseteq \starred{\flt}$.
\end{lemma}

Assigned qualifiers to well-typed values are lower bounds for observation in their typing.

\begin{lemma}[Values are Non-Fresh]
  If $\cx[\flt]{[\G\mid\Sigma]}\ts v:\ty[q]{T}$, then $\cx[\flt]{[\G\mid\Sigma]}\ts v:\ty[q\setminus\qfresh]{T}$.
\end{lemma}

In the mechanized proof, we define an invertible value typing that inlines subtyping, eliminating the need for handling subsumption induction cases.
These lemmas are then proved by induction on the invertible value typing.

\subsubsection{Well-Stepped Terms.}
The well-stepped terms allow us to only reason about the terms reduced from some statically well-typed terms.
The following lemmas are independent of the typing environment, so can be proved by induction over term reduction.

\begin{lemma}[Well-Stepped Values] \label{theorem:well-stepped-value}
  If $\WT{v}$, then $\LC{t}=\varnothing$.
\end{lemma}

A well-stepped term always reduces to a well-stepped term under a well-formed store.

\begin{lemma}[Well-Stepping of Terms]
  If $\WT{t}$, and $\cx[\flt]{[\varnothing\mid\Sigma]}\ts\sigma\mid \kappa$ for some $\flt,\Sigma,\kappa$, and $t\mid\sigma \to t'\mid\sigma'$, then $\WT{t'}$.
\end{lemma}

Syntactically correct but semantically ill-formed terms are excluded by the well-stepped term judgement, including the existence of scope eliminations in fully reduced values or not yet reduced terms.

\subsubsection{Local Locations.}
The dynamic typing rules enforce the separation of the local locations of a term and its qualifier. 
The lemma below is proved by induction over dynamic typing.

\begin{lemma}[Non-Escaping of Local Locations] \label{theorem:scope-local-location}
  If \phantom{n}$\cx[\flt]{[\G\mid\Sigma]}\ts t:\ty[q]{T}$, then $\killsep{\LC{t}}{q}$.
\end{lemma}

The combination of \Cref{theorem:scope-local-location}, \Cref{theorem:scope-value-retype}, and \Cref{theorem:well-stepped-value} reasons about the safe kill.
A well-stepped term with local locations reduces to a value with empty local locations, and can be retyped in a shrinking observation separate from its previous local locations before reduction.

\subsubsection{Substitution.}
In the call-by-value reduction, substitutions required by $\beta$-reduction are only considered for values containing no term variables, called top-level substitutions.
The substitution lemmas generally follow the proof tree of \citet{david}.

\begin{lemma}[Substitution Preserves Transitive Closures] \label{theorem:subst-preserve}
  If \phantom{n}$x:\ty[q]{T} \in \Gamma$, and $p,q\subseteq \starred{\DOML(\Sigma)}$, and $p\cap \starred{\flt}\subseteq q$, and $\qtrans{r}{\Gamma} \subseteq \starred{\flt}$, then $\qtrans{r\theta}{\Gamma\theta}\subseteq \qtrans{r}{\Gamma}\theta$, with substitution $\theta=[p/x]$. 
\end{lemma}
\begin{proof}
  By transitively applying subtyping, commutativity between transitive closures and substitutions, and monotonicity of qualifier substitution. In each iteration of transitive lookup on a variable, the qualifier substitution afterwards produces a qualifier no smaller than the substitution beforehand. 
\end{proof}

\begin{lemma}[Top-Level Substitution] \label{theorem:substitution}
  If \phantom{n}$\cx[\flt]{[x:\ty[q]{T},\G\mid\Sigma]}\ts t:Q$, and $\cx[p]{[\varnothing\mid\Sigma]}\ts v:\ty[p]{T}$, and $p\cap\starred{\flt}$, and $q=p\lor q=\starred{p\cap r}$, additionally $\killsep{\LC{t}}{p}$, then for the substitution $\theta = [p/x]$, $\cx[\flt\theta]{[\G\theta\mid\Sigma]}\ts t[v/x] : Q\theta$.
\end{lemma}
\begin{proof}
  By induction over the term typing of $t$.

  Specially in the case \rulename{t2appfr} and \rulename{t2tappfr}, applying the induction hypothesis requires $(p\overlap q)\theta \subseteq p\theta \overlap q\theta$ established from \Cref{theorem:subst-preserve}.
  In case \rulename{t2sub}, an analogous substitution lemma for subtyping and qualifier subtyping is required.

  The condition $\killsep{\LC{t}}{p}$ preserves that substitution cannot cause local locations escaping from their defined scope.
\end{proof}

\subsubsection{Main Soundness Result}

\begin{theorem}[Progress] \label{theorem:scope-progress}
  If \phantom{n}$\cx[\flt]{[\varnothing\mid\Sigma]}\ts t:Q$, and $\WF{\Sigma}$, and $\WT{t}$, then either $t$ is a value, or for any store $\sigma$ and killed locations $\kappa$ where $\cx[\flt]{[\varnothing\mid\Sigma]}\ts \sigma\mid\kappa$, there exists a term $t'$ and store $\sigma'$ such that $t\mid\sigma \to t'\mid\sigma'$.
\end{theorem}
\begin{proof}
  By induction over the term typing $\cx[\flt]{[\varnothing\mid\Sigma]}\ts t:Q$.
\end{proof}

\begin{theorem}[Preservation] \label{theorem:scope-preservation}
  If \phantom{n}$\cx[\flt]{[\varnothing\mid\Sigma]}\ts t:\ty[q]{T}$, and $\WF{\Sigma}$, and $\WT{t}$, and $t\mid\sigma \to t'\mid\sigma'$ for some $t'$ $\sigma'$, and $\cx[\flt]{[\varnothing\mid\Sigma]}\ts \sigma \mid \kappa$ for some $\kappa$, then there exists $\Sigma'$ $\flt'$ $\kappa'$ such that $\Sigma,\flt,\kappa\envto{t}\Sigma',\flt',\kappa'$, and $\cx[\flt']{[\varnothing\mid\Sigma']}\ts \sigma'\mid\kappa'$, and \(\cx[\flt']{[\varnothing\mid\Sigma']}\ts t': \ty[{q[p/\qfresh]}]{T}\) for \(p\subseteq \DOML(\Sigma'\setminus\Sigma)\).
\end{theorem}
\begin{proof}
  By induction over the typing $\cx[\flt]{[\varnothing\mid\Sigma]}\ts t:\ty[q]{T}$ and subsequently induction over the environment update $\Sigma,\flt,\kappa\envto{t}\Sigma',\flt',\kappa'$. 

  For case \rulename{t2app} with two sub-terms, the observation $\flt\ominus t_1$ is distinguished separately for reduction of $t_1$ or $t_2$:

  1. In the left congruence case of reduction on $t_1$, $t_2$ has not been reduced yet. The observation $\flt\ominus t_1$ grows monotonically with the permanent allocation during the reduction of $t_1$, but will not reflect the shrinking because all killed locations are already within the local locations of $t_1$.

  2. In the right congruence case of reduction on $t_2$, $t_1$ has already been reduced to a value. By \Cref{theorem:well-stepped-value}, $t_1$ has no local locations, so the observation of $t_2$ becomes $\flt$. The growth of observation by reduction on $t_2$ does not reflect on the typing of $t_1$ by \Cref{theorem:scope-value-observation}.

  3. In the contraction case where $t_1$ and $t_2$ are both values, all the local locations have already been killed by \Cref{theorem:well-stepped-value}. The substitution \Cref{theorem:substitution} is applied twice for argument $x$ and self-reference $f$ to establish the case.
  
  Other cases with two sub-terms handle local locations similarly. The contraction case of scope elimination $\twithc{\ell}{v}$ completes the proof by applying \Cref{theorem:scope-value-retype} to shrink the observation.
\end{proof}

All results are mechanized in Rocq; we refer readers to our artifact for details. %
\section{Case Studies}  \label{sec:casestudy}

In this section, we demonstrate expressiveness through three case studies: 
relaxation of telescoping structures enabling a general fixed-point combinator without cyclic references (\Cref{sec:case-fixpoint});
non-lexical arenas for resource management (\Cref{sec:case-callback}); 
and safe construction and reclamation of cyclic store structures with multi-hop cycles (\Cref{sec:case-multihop}).

\subsection{General Fixed-Point Combinator} \label{sec:case-fixpoint}

A general fixed-point combinator is typically implemented by storing a function in a mutable reference that refers to itself. 
\citet{david} model this with cyclic references: the referent can observe the reference itself, permitting self-cycles in the store topology.

Such cyclic references impose extra constraints. In particular, the reference's outer qualifier must remain fixed to avoid enlarging the referent's reachability, which would violate telescoping well-formedness. 
Consequently, \citet{david} restrict deep substitution\footnote{See discussion in Section 6, \citet{david}. Compared with rules \rulefmt{t-app} and \rulefmt{t-app\qfresh} in \citet{wei24}.} of function arguments to empty or singleton qualifiers, limiting expressiveness.

By contrast, our growable arenas with coarse-grained reachability permit recursive patterns without cyclic references, avoiding this limitation:
\begin{lstlisting}
  val a = new Ref()
  def fix[T] (f: (g: (T -> T)$\trackvar{a}$ -> (T -> T)$\trackvar{g}$)) : (T -> T)$\trackvar{a}$ = {
    val c = new Ref(x => x) at a        // : Ref[(T -> T)$\trackvar{a}$]$\trackvar{c}$
    c := f((n : T) => (!c)(n))          // argument : (T -> T)$\trackvar{c}$ <: (T -> T)$\trackvar{a}$
    !c                                  // : (T -> T)$\trackvar{a}$
  }
\end{lstlisting}
Instead of allocating a fresh reference per recursive call, the function is stored in @c@, which is coallocated within arena~@a@. 
Since all objects in @a@ share a common reachability identity, internal self-references such as @c@ are permitted without violating telescoping constraints.
We implement the fixed-point operator and a factorial example in the \arenalang-calculus (with a trivial integer extension); we refer readers to our artifact for details.

\subsection{Callback Registration} \label{sec:case-callback}

In event-driven systems, callbacks enable asynchronous programming and flexible control flow. 
While typically registered during initialization, callbacks may be invoked long after registration and outside their defining scope.
Delayed invocation requires callback resources to remain alive until explicit deregistration. To achieve this, callbacks are stored in a non-lexical resource pool that outlives their defining scope and is encapsulated by the registration closure to prevent misuse.

Non-lexical arenas with coarse-grained reachability satisfy these requirements. 
Handlers are coallocated in a single arena captured by the registration closure, which remains opaque to users. 
We sketch the callback registration API as follows:
\begin{lstlisting}
  val makeHandler = {
    val rp = new Ref(())                // non-lexical resource pool
    (cb: Int => Unit) => {
      val h = new Ref(cb) at rp         // : Ref[(Int => Unit)]$\trackvar{h}$
      h                                 // return handler                       
    }                                   // : ((cb: Int => Unit) => Ref[Int => Unit]$\trackvar{rp}$)$\trackvar{rp}$
  }                                 
  val f = (x: Int) => ()                // : (Int => Unit)$\trackvar{f}$
  makeHandler(f)                        // : Ref[Int => Unit]$\trackvar{makeHandler}$
\end{lstlisting}
The arena @rp@ is a shared pool for registered callbacks.
Each call to @makeHandler@ returns a handler coallocated in @rp@ and encapsulated by the closure. 
Crucially, callbacks are tracked via reachability to @makeHandler@, and deallocating @makeHandler@ reclaims all handlers in bulk.

\subsection{Circular Store Structures} \label{sec:case-multihop}

Reference counting is a popular memory management technique in which each object maintains a count of active references pointing to it.
When the count drops to zero, the object is automatically reclaimed.
Though simple and efficient in many cases, reference counting fails to reclaim objects forming multi-hop cycles where none of the reference counts ever reaches zero.

Our system properly handles both the construction and safe deallocation of such cyclic structures: %
\begin{lstlisting}
  { // scope starts
    val a = new Ref(()) scoped          // : Ref[Unit]$\trackvar{a}$
    val f = (x : Int) => 7              // : (Int => Int)$\trackvar{f}$ <: (Int => Int)$\trackvar{a}$
    val c1, c2, c3 = new Ref(f) at a    // : Ref[(Int => Int)$\trackvar{a}$]$\trackvar{..}$
    c1 := x => { (!c2)(x) }             // OK, Ref[(Int => Int)$\trackvar{a}$]$\trackvar{c2}$ <: Ref[(Int => Int)$\trackvar{a}$]$\trackvar{a}$
    c2 := x => { (!c3)(x) }             // OK
    c3 := x => { (!c1)(x) }             // OK
  } // deallocate {a, r1, r2, r3}
\end{lstlisting}
The references @c1@, @c2@, and @c3@ form a transitive multi-hop cycle through the store.
Because reachability is tracked at the arena level, the three references (@c1@, @c2@, and @c3@) coallocated with the proxy reference~(@a@) are grouped in a single reachability unit. 
The entire structure behaves as a circular linked list and is treated as a whole under coarse-grained reachability tracking. 
At scope end, the scoped reference @a@ and its coallocated references are deallocated in bulk, avoiding the pitfalls of reference counting while preserving timely and safe reclamation.

Circular patterns, where a reference must transitively reach previously allocated intermediaries, break telescoping structures and lie beyond prior reachability type systems \cite{david,wei24}. 
Our work addresses this challenge.
Moreover, this design can generalize to doubly linked lists, which can be implemented via Church encoded pairs \cite{wei24} storing two closures within each node.

\section{Discussion and Limitations}  \label{sec:discussion}

\paragraph{Imprecision from Flow-Insensitivity.}
Our approach is type-based, meaning that reachability types and our deallocation reasoning atop are flow-insensitive, which in general strikes a good balance between precision and complexity like most type systems.
Flow-insensitivity in our design arises from two sources: 
(1) reachability qualifiers themselves, which over-approximate reachable resources; and 
(2) deallocation reasoning based on the local locations over reduction (\Cref{sec:scope-reasoning}), which does not introduce extra imprecision.

Although our deallocation reasoning is lightweight in annotation, it is less expressive than flow-sensitive effect systems as informally described in \citet{wei24} and more generally by \citet{DBLP:journals/toplas/Gordon21}.
Specifically, deallocation is guaranteed only for scoped arenas and at scope exit, not for arbitrary arenas or in the middle of a scope.
We regard flow-sensitive studies as parallel work, including recent representative efforts by \citet{deng2025freemovereachabilitytypes} and \citet{jia2025typestaterevocablecapabilities}.

\paragraph{Garbage Collection.}
Our system focuses on impure high-level languages where the choice of using garbage collection (GC) is common.
We do not aim to remove GC entirely; instead, we provide the additional benefits of timely guaranteed reclamation via selective stack discipline when desired.
This mechanism can ultimately apply to resources beyond reference cells, such as files or sockets.
Our language design is also open to having internal GC within an arena as a complementary option.

Completely eliminating GC would be possible but too restrictive because it forbids all non-lexical arenas and escaping patterns.
This is consistent with Rust's use of reference counting: although Rust's ownership system does not rely on GC, adding reference counting allows additional expressiveness such as sharing.

\paragraph{Type Inference and Empirical Implementation.}
Sound, expressive, and efficient
type checking and inference remain open challenges for reachability types. %
In particular, typing an escaped term (\Cref{sec:overview-RTrecap}, \Cref{sec:overview-nonlexical-escaping}) is subject to the avoidance problem \cite{jia2025escapeselfsoundexpressive}: 
a variable mentioned in types becomes ill-scoped when its defining scope ends
and thus needs replacement.
In this work and prior reachability type systems,
such replacement requires explicit term-level coercions (omitted for clarity),
rather than a more natural way to upcast via subtyping. 
Improving the subtyping and deriving typing algorithms are important but orthogonal to this work.

\paragraph{Data Structures.}
In this work, we have shown how to encode cyclic store structures with multi-hop cycles, a key step towards supporting practical cyclic data structures.
Similar goals have appeared in related works: \citet{DBLP:journals/pacmpl/XuBO24} demonstrates that capturing types can model mutable, cyclic object graphs via class extensions, though they stop at providing the full details of the class extensions.
We likewise leave a complete object encoding to future work.

\section{Related Work} \label{sec:related-word}

\paragraph{Regions/Arenas.} 
{Early} forms of regions/arenas were explored under various names with explicit management,
but without static safety guarantees.
\citet{arena_single} introduced \emph{arenas}
for fast allocation and deallocation
by manually grouping objects by lifetime.
\citet{region_local_reasoning} explored a similar idea for ease of
reasoning, called \emph{local stores}.
\citet{explicit_region} identified safety limitations in these systems 
and introduced a mechanism that skips deallocation for any region still referenced.
Their runtime approach demonstrated performance advantages, especially for list structures,
despite incurring reference counting overhead.
A follow-up study \cite{region_language_support} explored annotating structural information
to mitigate the cost of reference counting.

Prominent examples of region-based systems include ML Kit \cite{region-original,MLkit_new}
and Cyclone \cite{cyclone}. Both enforce static safety but differ in languages and styles:
ML Kit targets on a functional language with \emph{implicit} region management via
a Hindley-Milner style inference, while
Cyclone adopts \emph{explicit} regions in a low-level C-like language
augmented with existential types to encode closures.
In this work, arenas require explicit allocation and scoping, but unlike traditional explicit
systems, regions keys or handles are not needed---existing references suffice.
This allows region polymorphism to be naturally expressed as function abstraction.

Beyond surface languages, MLKit \cite{MLkit_new} and Cyclone \cite{cyclone}
share foundational elements.
Both rely on forms of effect (liveness) types for static guarantees,
and following the insight \cite{DBLP:journals/tocl/WadlerT03} that effects
can be transposed as monads, both can be embedded in a monadic region calculus \cite{monadic_regions}.
While lexically scoped lifetime have been noted as a limitation,
both systems provide dynamically managed global regions to compensate,
although this sacrifices some of the performance and isolation benefits of regions.
Further development in Cyclone \cite{fluetImplementationPerformanceEvaluation,DBLP:conf/iwmm/HicksMGJ04}
introduced escaping arenas that require additional capability management.
These extensions were later formalized and proven sound via linear regions \cite{first_class_region}.

\paragraph{Region Logic} 
In assertion-based verification techniques,  
region logic~\cite{region_logic} allows users to declare and update ghost fields and variables of type ``region''.
Regions in their work store objects, and have been refined to storing memory locations~\cite{relational_logic_effect,DBLP:journals/fac/BaoLE18}. 
In our work, the reachability qualifier of each arena serves a similar purpose: it provides an upper bound on the locations that may store values belonging to the arena.
We envision that our approach could be leveraged to synthesize those statements (in region logic) through reachability qualifiers, 
thereby, reducing annotation overhead in their frameworks.
We leave a proper investigation as future work.

\paragraph{Linear Types, Ownership Types, and Their Region Variants.}
Beyond regions, other mechanisms for static lifetime control exist in flow-sensitive
forms. Linear type systems \cite{wadler1990linear, DBLP:conf/fpca/TurnerWM95}
require each variable to be used exactly once. While this prevents sharing, it
streamlines reasoning about lifetime and deallocation.
Ownership types \cite{DBLP:conf/oopsla/ClarkePN98,DBLP:conf/ecoop/NobleVP98,DBLP:series/lncs/ClarkeOSW13}
are often designed in object-oriented languages, where
variants differ in their topological restrictions and encapsulation disciplines.
Rust \cite{rust} is a notable instance with growing adoption, enforcing unique ownership and a 
single mutable access path throughout execution.
This model enables predictable memory deallocation.
Meanwhile, such strict invariants 
can pose challenges for users \cite{DBLP:journals/pacmpl/CrichtonGK23} and
complicate the safe construction of cyclic data structures
\cite{fearless, DBLP:journals/pacmpl/YanovskiDJD21}.
Proposals have been made for relaxing the uniqueness requirements in Rust
\cite{rusty_links} and ownership types \cite{DBLP:conf/oopsla/MullerR07}.

Efforts to integrate regions with linear or ownership types aim to
support both flexible sharing and non-lexical lifetime control.
Linear regions \cite{regions_linear_type,first_class_region} track regions
and resources in unrestricted ways, whereas region capabilities are
tracked linearly, required for granting access and consumed by deallocation.
Pony \cite{pony} employs implicit regions for actor programming,
decomposing fractional capabilities for fine-grained control.
Verona's region system, Reggio \cite{verona}, combines ownership with explicit regions
to ensure isolation among threads in concurrent programming:
each region can be managed by a configured strategy, and
each thread operates within a window of a single mutable region.
These systems generally require users to explicitly \emph{enter} or \emph{focus} on a region to
obtain the necessary capabilities---a hallmark of flow-sensitive reasoning.
\citet{fearless} formalized these mechanisms as \emph{virtual transformations}
and proposed an inference scheme to reduce the manual overhead.

\paragraph{Reachability Types.}

This work extends reachability types with new mechanisms for 
store management and scoping.
Seeking to adapt separation logic \cite{separation_logic} to functional languages,
\citet{bao21} introduced reachability qualifiers,
sets of variables annotated on types, to express resource access patterns.
\citet{wei24} refined this idea by incorporating \emph{one-step} reachability and
freshness, yielding a more precise and dynamic model.
Their polymorphic calculus~$\weipolylang$ tracks resource sharing without enforcing
global invariants, offering a flexible foundation for reasoning about resources
in higher-order settings. 
Graph IR \cite{10.1145/3622813} used reachability types to optimize impure higher-order programs.
We build upon this framework to develop our calculi.

The semantics of the monomorphic $\weilang$ were analyzed via logical relations
\cite{10.1145/3763116},
and, counterintuitively, proven terminating despite its extension with a higher-order store.
This result stems from the store's adherence to the telescoping requirement
(see \Cref{sec:overview-telescope}), which further motivated efforts to enable
cycles in the store \cite{david}.
Their \cycliclang allows self-cycles (see \Cref{fig:intro-topo-rt}) and
thus non-terminating computation (\eg, fixed-point combinators),
but imposes extra constraints on application rules to do so.
By contrast, our $\arenalang$ allows similar patterns via intra-arena cycles
without such constraints, although cycles require at least two cells.
\citet{david} further proposed refinements to the reference model, which we incorporate in our design.

Reachability types are inherently flow-insensitive and thus cannot
express linearity/uniqueness in isolation.
Flow-sensitive effect extensions have been discussed regarding this limitation \cite{bao21,wei24},
including destructive effects for deallocation.
These approaches would require additional effect qualifiers
and corresponding reasoning in typing judgements.
Our $\withlang$, in contrast, ensures scoped deallocation safety by
parsing lifetime information directly from reachability qualifiers.
To support explicit, flow-sensitive deallocation (\eg, manually calling @free@),
a separate flow-sensitive effect extension would still be required.

With a similar design to reachability types,
capturing types \cite{DBLP:journals/toplas/BoruchGruszeckiOLLB23,DBLP:journals/pacmpl/XuBO24} were initially
proposed for effect safety in Scala \cite{DBLP:conf/scala/OderskyBBLL21,DBLP:journals/corr/abs-2207-03402}.
Their design allows to work around the telescoping requirement
(see \Cref{sec:overview-telescope})
by employing a top qualifier @{cap}@. Top qualifiers, however,
similar to \emph{top types}, offer little information in reasoning,
as a resource tracked as @{cap}@ can potentially reach/capture anything.

\section{Conclusion} \label{sec:conclusion}

In this work, we unified reachability types, arena-based resource management, and stack discipline.
We introduced the \arenalang-calculus, which generalizes reachability reasoning to non-lexical shadow arenas over a two-dimensional store model.
By lifting reachability tracking from fine-grained to coarse-grained, \arenalang relaxes the telescoping structures that limit cyclic store structures.
Building upon \arenalang-calculus, the novel \withlang-calculus reestablishes selective stack discipline through scoped allocation, enabling sound, bulk, and flow-insensitive deallocation reasoning.
Both calculi were formalized and proven type sound and memory safe in Rocq.
This unified framework provides a lightweight yet expressive foundation for safe resource management in higher-order languages, reconciling first-class arenas and scoped lifetime control within a single type system.

\section*{Data Availability Statement}
Rocq mechanizations can be found at \url{https://github.com/tiarkrompf/reachability} .

\begin{acks}                            %
  We thank Haotian Deng for related contributions
  to reachability types.
  This work was supported in part by NSF award 2348334 and Augusta faculty startup
  package, as well as gifts from Meta, Google, Microsoft, and VMware.
\end{acks}

\bibliography{references}

\end{document}